\documentclass[12pt]{article} 

\usepackage{epsfig}
\usepackage{graphicx}
\usepackage{comment}
\usepackage{latexsym}
\usepackage{hyperref}
\usepackage{amsmath}
\usepackage{color}
\usepackage{amsbsy}
\usepackage{amssymb}
\usepackage{amsthm}
\usepackage{amsfonts}
\usepackage{cite}

\newcommand{\arXivold}[1]{\href{http://arxiv.org/pdf/#1}{{\tt [#1]}}}

\newcommand{\nF}{n_{\rm F}}
\newcommand{\nB}{n_{\rm B}}
\newcommand{\F}{{\cal F}}
\newcommand{\N}{{\cal N}}
\newcommand{\V}{{\cal V}}
\newcommand{\D}{{\cal D}}

\newcommand{\G}{\boldsymbol{G}}

\newcommand{\be}[0]{\begin{equation}}
\newcommand{\ee}[0]{\end{equation}}

\renewcommand{\ni}{\noindent}

\newcommand{\R}{\mathbb{R}}
\newcommand{\Z}{\mathbb{Z}}

\renewcommand{\O}{{\cal O}}
\renewcommand{\Re}{{\rm Re}\,}
\renewcommand{\Im}{{\rm Im}\,}

\newcommand{\Str}{\textrm{Str}\,}

\newcommand{\ie}{{\em i.e.} }

\newcommand{\via}{{\it via} }

\newcommand{\where}{\mbox{where}}

\newcommand{\when}{\mbox{when}}
\renewcommand{\and}{\mbox{and}}

\newcommand{\esp}{\phantom{\!\!\overset{\displaystyle |}{|}}}
\newcommand{\espD}{\phantom{\!\!\underset{\displaystyle |}{\cdot}}}
\newcommand{\espDD}{\phantom{\!\!\underset{\displaystyle |}{|}}}

\newcommand{\bm}{\boldmath} 

\catcode`\@=11
\def\marginnote#1{}
\newcount\hour
\newcount\minute
\newtoks\amorpm
\hour=\time\divide\hour by60 \minute=\time{\multiply\hour by60
\global\advance\minute by-\hour}
\edef\standardtime{{\ifnum\hour<12 \global\amorpm={am}%
        \else\global\amorpm={pm}\advance\hour by-12 \fi
        \ifnum\hour=0 \hour=12 \fi
        \number\hour:\ifnum\minute<10 0\fi\number\minute\the\amorpm}}
\edef\militarytime{\number\hour:\ifnum\minute<10 0\fi\number\minute}
\def\draftlabel#1{{\@bsphack\if@filesw {\let\thepage\relax
   \xdef\@gtempa{\write\@auxout{\string
      \newlabel{#1}{{\@currentlabel}{\thepage}}}}}\@gtempa
   \if@nobreak \ifvmode\nobreak\fi\fi\fi\@esphack}
        \gdef\@eqnlabel{#1}}
\def\@eqnlabel{}
\def\@vacuum{}
\def\draftmarginnote#1{\marginpar{\raggedright\scriptsize\tt#1}}
\def\draft{\oddsidemargin -.2truein
        \def\@oddfoot{\sl preliminary draft \hfil
        \rm\thepage\hfil\sl\today\quad\militarytime}
        \let\@evenfoot\@oddfoot \overfullrule 3pt
        \let\label=\draftlabel
        \let\marginnote=\draftmarginnote
   \def\@eqnnum{(\theequation)\rlap{\kern\marginparsep\tt\@eqnlabel}%
\global\let\@eqnlabel\@vacuum}  }
\def\thebibliography#1{
\vskip 0.5cm 
\noindent
{\bf \Large References}
\list{
[\arabic{enumi}]}{\settowidth\labelwidth{[#1]}
\leftmargin\labelwidth
\advance\leftmargin\labelsep
\usecounter{enumi}}
\def\newblock{\hskip .11em plus .33em minus .07em}
\sloppy\clubpenalty4000\widowpenalty4000
\sfcode`\.=1000\relax}

\renewcommand{\theequation}{\arabic{section}.\arabic{equation}}
\renewcommand{\section}{\setcounter{equation}{0}\@startsection
{section}{1}{0mm}{-\baselineskip}{0.5\baselineskip} {\normalfont\Large\bfseries}}
\renewcommand{\subsection}{\@startsection
{subsection}{2}{0mm}{-0.6\baselineskip}{0.1\baselineskip} {\normalfont\large\bfseries}}

\renewcommand{\subsubsection}{\@startsection
{subsubsection}{3}{0mm}{-\baselineskip}{0.5\baselineskip}
{\normalfont\normalsize\slshape}}

\begin{document}


\begin{titlepage}
\begin{flushright}
CPHT-RR119.122018, December   2018
\vspace{1cm}
\end{flushright}
\begin{centering}
{\bm\bf \Huge The Casimir Effect in String Theory}

\vspace{5mm}

 {\bf Alexandros Kehagias$^{1}$ and Herv\'e Partouche$^2$}

 \vspace{3mm}

$^1$ Physics Division, National Technical University of Athens,\\
15780 Zografou Campus, Athens, Greece\\
{\em  kehagias@central.ntua.gr}

$^2$  {Centre de Physique Th\'eorique$^\dag$, Ecole Polytechnique, \\
F--91128 Palaiseau cedex, France\\
{\em herve.partouche@polytechnique.edu}}

\end{centering}
\vspace{0.7cm}
$~$\\
\centerline{\bf\Large Abstract}\\

\begin{quote}

We discuss the Casimir effect in heterotic string theory. This is done by considering a $\Z_2$ twist acting on one external compact direction and three internal coordinates. The hyperplanes fixed by the orbifold generator $\G$ realize the two infinite parallel plates. For the latter to behave as ``conducting material'', we implement in a modular invariant way  the  projection $(\boldsymbol{1}-\G)/2$ on the spectrum running in the  vacuum-to-vacuum amplitude at one-loop. Hence, the relevant projector to account for the Casimir effect is orthogonal to that commonly used in string orbifold models, which is $(\boldsymbol{1}+\G)/2$. We find that this setup yields the same net force acting on the plates in the context of quantum field theory and string theory. However, when supersymmetry is not present from the onset, finiteness of the resultant force in field theory is reached by adding formally infinite forces acting on either side of each plate, while in string theory both contributions are finite. On the contrary, when supersymmetry is spontaneously broken \`a la Scherk-Schwarz, finiteness of each contribution is fulfilled in field and string theory.

\end{quote}
\noindent
{\it Keywords:} Vacuum Energy, Casimir Force, Partition Function, Supersymmetry, Supersymmetry Breaking.
\vskip.1in
\noindent
{\it PACS numbers:} 11.25.-w, 11.10.-z, 11.30.Pb, 11.30.Qc
\vspace{3pt} \vfill \hrule width 6.7cm \vskip.1mm{\small \small \small
  \noindent
   $^\dag$\ Unit\' e  mixte du CNRS et de l'Ecole Polytechnique, UMR 7644.}\\

\end{titlepage}
\newpage
\setcounter{footnote}{0}
\renewcommand{\thefootnote}{\arabic{footnote}}
 \setlength{\baselineskip}{.7cm} \setlength{\parskip}{.2cm}

\setcounter{section}{0}


\section{Introduction}
The prototypical form of the Casimir force is the attraction felt by  two flat, closely spaced, parallel mirrors originating from the 
vacuum energy of the electromagnetic field between the plates 
\cite{rev,Lam0}. Precise measurements of the Casimir force become technically possible only about twenty years ago thanks to  the pioneering  experimental work of Lamoreaux \cite{Lam1}. It is also interesting to notice  that it took almost half a century 
for the first prediction of this force to be measured at the level of per cent accuracy. The force $F$ exerted between two parallel conducting plates of area $A$ and at a distance $L$ turns out to satisfy 
\begin{eqnarray}
\label{CFo}
\F\equiv \frac{F}{A}=-\frac{\hbar c \pi^2}{240 L^4}.
\end{eqnarray}
This force arises due to the structure of the  electromagnetic modes (with zero-point energy $\hbar \omega/2$ each)  between the two plates, as compared to free space without the plates. The reason is that in quantum field theory,  in the absence of gravity, only differences in energy have a physical meaning. Therefore, one should compare the energy of the two-plates configuration with some reference background, which is flat empty Minkowski space here. Of course, things are quite different when gravity is turned on. In this case, vacuum energy contributes to the energy-momentum tensor 
as a cosmological constant and  therefore to the spacetime curvature through Einstein equations. In other words, vacuum energy gravitates as all forms of energy.  However,  the cosmological constant turns out to be infinite in quantum field theory  and some regularization is needed, which shifts  the  cosmological constant to the cutoff scale. In string theory, on the other hand, there are no infinities in the calculation of the vacuum energy and therefore no need for regularization. The cosmological constant  is of the order of the supersymmetry breaking scale, however in conflict with observations. 
 
Nevertheless, the Casimir effect is  a remarkable result as it was the first indication of the vacuum energy, although it can be formulated and computed with no reference to any zero-point energy. 
In the latter approach, the Casimir force is  regarded as the retarded relativistic van der Waals force between the metal plates \cite{jaffe}.  In other words, this point of view suggests that the Casimir effect does not really support   the existence of vacuum energy of  quantum fields more than any other one-loop effect in quantum field theory. In that case, the Casimir force should vanish as the fine structure constant goes to zero, as all one-loop effects in quantum electrodynamics \cite{jaffe}. However, here we will follow the traditional  route and calculate the Casimir force by summing up zero-point energies in field and string theory in spacetime dimension $d$.  

To describe a universe comprised between two hyperplanes, we consider spacetime to be $\R^{1,d-1}\times S^1(R_{d-1})/\Z_2$, where the $\Z_2$ generator $\G$ acts as a twist $x^{d-1}\to -x^{d-1}$ on the circle of radius $R_{d-1}$. As a result, infinite parallel plates are located at the fixed points $x^{d-1}=0$ and $x^{d-1}=\pi R_{d-1}$. However, this setup is not quite what we are interested in for describing the Casimir effect. The reason is that the plates should be ``conducting'', which means that all modes propagating between the hyperplanes should have nodes at $x^{d-1}=0$ and $x^{d-1}=\pi R_{d-1}$. However, all degrees of freedom in the  background above are {\em even} rather than {\em odd} under the twist. If the orbifold projector $(\boldsymbol{1}+\G)/2$ is common (at least in string theory) to project the spectrum on even modes, it turns out that the correct orbifold projector for restricting to odd modes is the orthogonal one, $(\boldsymbol{1}-\G)/2$. 

The above result is shown in quantum field theory in Sect.~\ref{QFT}. Actually, the force acting on the plates is equal, whether even or odd modes are selected, as the vacuum amplitude with $\G$ inserted projects on modes non-propagating along $x^{d-1}$ and thus not contributing to the transverse force. We analyse the case where supersymmetry is not present from the onset and recover the fact that the Casimir force can eventually be interpreted as the finite resultant of two formally infinite forces exerted on either side of each plate: One derived from the vacuum between the hyperplanes, and one derived  from the vacuum in an infinite half space. However, we show that when supersymmetry  is present in a spontaneously broken phase, these forces are individually finite.  

In Sect.~\ref{Orbi}, we generalize the orbifold prescription to the heterotic string theory. For consistency arising from modular invariance, $\G$ also acts  on three internal directions we take toroidal. Moreover, a twisted sector arises, which has no counterpart in field theory. However, the twisted spectrum being localized on the fixed hyperplanes, it does not contribute to the transverse force. Therefore, string theory reproduces in great details the field theory derivation of the forces acting of either side of each plate, which are individually finite when supersymmetry is spontaneously broken. The outcomes in both frameworks match up to contributions arising from heavy string modes, which turn out to be irrelevant as they are exponentially suppressed. 

In Sect~\ref{hb}, we consider the $SO(16)\times SO(16)$ heterotic string~\cite{o16}, which is known to be explicitly non-supersymmetric. In this case, finiteness of each individual force remains valid, due to the natural regularization of the vacuum amplitude arising from the infinite spectrum with unbounded masses. In other words, when supersymmetry is broken in a hard way, not only the heavy states contribute significantly, since they are even rendering the vacuum energy finite in every case, whether the plates are at finite or infinite distance. This is the great advantage of string theory over field theory where differences of energies can only be considered, for the infinite sums over zero-point energies to cancel out. However, considering the energy gap between the configurations where the plates are either at finite distance, or infinite distance, the field theory and string theory outcomes are in perfect agreement, up to exponentially suppressed corrections arising from heavy string states. 

Our conclusion can be found in Sect.~\ref{conclus}, and we mention that 
throughout this work, we use string units, $\alpha'=1$, and the notations of Ref.~\cite{kiritsis}.

\section{Casimir effect in quantum field theory}
\label{QFT}


In this section, our aim is to review and then develop different approaches for computing the Casimir force in field theory, when supersymmetry is not present from the onset, or when it is realized in a spontaneously broken phase. In particular, we will introduce an orbifold point of view suitable for later generalization in the framework of string theory. The field  theory analysis will be presented in $d$ dimensions for a real scalar field, or for a single degree of freedom, fermionic or bosonic. 


\subsection{Second quantized formalism}
\label{secondQ}

Our basic consideration is the path integral computation of the partition function of a real bosonic free field. In dimension $d$ and for a mass $M_{\rm B}$, the partition function is 
\be
\begin{aligned}
Z_{\rm B}&=\int\D\phi \; e^{ -i {\textstyle \int} d^dx\, {1\over 2} (\partial_\mu\phi\partial^\mu\phi+M_{\rm B}^2\phi^2)}\\
&=\int\D\phi \; e^{-{\textstyle \int }d^dx_E \, {1\over 2}\, \phi(-\partial_\mu\partial_\mu+M_{\rm B}^2)\phi}.
\end{aligned}
\ee  
The former expression uses Lorentzian metric $(-1,1,\dots,1)$, while the latter is obtained by Wick rotation $x_E^0=-ix^0$. We use a compact version $\mathbb{E}^d$ of the Euclidean spacetime, with finite volume $V_{0,\dots,d-1}$,  
\be
\mathbb{E}^d=\prod_{\mu=0}^{d-1}S^1(R_\mu), 
\qquad V_{0,\dots,d-1}=\prod_{\mu=0}^{d-1}2\pi R_\mu,
\ee
where all directions are circles of radii $R_\mu$.
Imposing the field to have periodic boundary conditions, its mode expansion in orthonormal modes satisfies
\be
 \phi(x_E)={1\over \sqrt{V_{0,\dots,d-1}}}\sum_{m\in \Z^d}\phi_m\, e^{i{m_\mu\over R_\mu}x^\mu_E}, \qquad \phi_{-m}=\phi_m^*.
 \label{modesb}
\ee
By using the above expression, the path integral can be written as 
\be
\begin{aligned}
Z_{\rm B}&=  \int\D\phi\, e^{\textstyle-{1\over 2} \sum_{m\in\Z^d} \big[({m_\mu\over R_\mu})^2+M_{\rm B}^2)\big]|\phi_m|^2}\\
&=\int{d\phi_0\over \sqrt{2\pi}}\, e^{-{1\over 2}M_{\rm B}^2\phi_0^2}\prod_{m\in\Z^d_+}\int {d\Re\phi_m \over \sqrt{\pi}}\, {d\Im\phi_m\over \sqrt{\pi}}\, e^{{\textstyle -\big[({m_\mu\over R_\mu})^2+M_{\rm B}^2\big]}\!\left[(\Re\phi_m)^2+(\Im\phi_m)^2\right]}, \label{zbb}
\end{aligned}
\ee
where $\Z^d_+$ is the set of non-vanishing $d$-tuples whose first nonzero entry is positive. The integration over the real and imaginary parts of the  modes $\phi_m$  being Gaussian, we find the well know result
\be
\ln Z_{\rm B}=-{1\over 2}\sum_{m\in \Z^d}\ln\! \Big[\Big({m_\mu\over R_\mu}\Big)^2+M_{\rm B}^2\Big].
\label{pi}
\ee

\vskip.06in
\noindent
{\bf Splitting even and odd modes:}
In Minkowski spacetime, Casimir effect arises between two parallel, infinite hyperplanes, we choose to be located at $x^{d-1}=0$ and $x^{d-1}=\pi R_{d-1}$. A field between the hyperplanes has nodes on the ``conducting'' boundary plates. To make contact with the unconstrained field $\phi$ periodic along $x^{d-1}$, we define
\be
\begin{aligned}
\forall \hat x_E\in\R^{d-1}, \;\;\forall x^{d-1}\in \R, \qquad &\phi^{\rm e}(\hat x_E,x^{d-1})={1\over 2}\big(\phi(\hat x_E,x^{d-1})+\phi(\hat x_E,-x^{d-1})\big),\\
&\phi^{\rm o}(\hat x_E,x^{d-1})={1\over 2}\big(\phi(\hat x_E,x^{d-1})-\phi(\hat x_E,-x^{d-1})\big),
\end{aligned}
\ee
so that $\phi=\phi^{\rm e}+\phi^{\rm o}$, where  $\phi^{\rm e}$, $\phi^{\rm o}$ are respectively even and odd under $x^{d-1}\to -x^{d-1}$ and  $2\pi R_{d-1}$-periodic. Actually, $\phi^{\rm e}$ contains all modes $\cos(\pi x^{d-1}/R_{d-1})$, $m_{d-1}\ge 0$, while  
 $\phi^{\rm o}$ contains all modes $\sin(\pi x^{d-1}/R_{d-1})$, $m_{d-1}\ge 1$. 
 
 We can  now define  path integrals $Z_{\rm B}^{\rm e},
 ~Z_{\rm B}^{\rm o}$ for only even  and odd fields, respectively as 
 \be
  Z_{\rm B}^{\rm e} =\int\D\phi^{\rm e} \; e^{-{\textstyle \int }d^dx_E \, {1\over 2}\, \phi^{\rm e}(-\partial_\mu\partial_\mu+M_{\rm B}^2)\phi^{\rm e}},\quad Z_{\rm B}^{\rm o} =\int\D\phi^{\rm o} \; e^{-{\textstyle \int }d^dx_E \, {1\over 2}\, \phi^{\rm o}(-\partial_\mu\partial_\mu+M_{\rm B}^2)\phi^{\rm o}},
 \ee
where the integrals in the actions cover the whole circle,\footnote{We may choose to integrate only over $0\le x^{d-1}\le \pi R_{d-1}$. This is a matter of convention, since this may be compensated by a rescaling of the fields and path integral measure.} $-\pi R_{d-1}\le x^{d-1}\le \pi R_{d-1}$. In these notations, the partition function in Eq.(\ref{zbb}) can be written as 
 \be
 Z_{\rm B}=\int\D\phi^{\rm e}\D\phi^{\rm o} \; e^{-{\textstyle \int }d^dx_E \, {1\over 2}\, (\phi^{\rm e}+\phi^{\rm o})(-\partial_\mu\partial_\mu+M_{\rm B}^2)(\phi^{\rm e}+\phi^{\rm o})}= Z_{\rm B}^{\rm e} Z_{\rm B}^{\rm o},
\ee
where the contributions of the cross products in the action vanish due to the parity of the fields. In fact, it is straightforward to verify that
\be
\begin{aligned}
\ln Z_{\rm B}^{\rm e}=-{1\over 2}\sum_{m\in \Z^{d-1}}\sum_{m_{d-1}\ge 0}\ln\! \Big[\Big({m_\mu\over R_\mu}\Big)^2+M_{\rm B}^2\Big],\\ 
\ln Z_{\rm B}^{\rm o}=-{1\over 2}\sum_{m\in \Z^{d-1}}\sum_{m_{d-1}\ge 1}\ln\! \Big[\Big({m_\mu\over R_\mu}\Big)^2+M_{\rm B}^2\Big].
\label{pi2}
\end{aligned}
\ee
Several problems can then be analyzed: 

$\bullet$ We can keep the whole set of modes $m\in\Z^d$, in order to describe a universe circular along $x^{d-1}$.

$\bullet$ We can also restrict to the cosine modes $m_{d-1}\ge 0$, to describe a universe with an orbifolded direction $S^1(R_{d-1})/\Z_2$, where the $\Z_2$ generator $\G$ acts as $x^{d-1}\to -x^{d-1}$ and where all fields are {\em even}. 

$\bullet$ Finally, we can restrict to the sine modes $m_{d-1}\ge 1$, in order to describe a ``Casimir-like universe'', \ie with orbifolded direction but with all fields  {\em odd} under the generator $\G$, \ie with nodes on the hyperplanes. 

To express these remarks in a more formal way, we can associate to the modes used in the expansion~(\ref{modesb}) the orthonormal basis of ``bra'' $|m\rangle\equiv |m_0,\dots,m_{d-1}\rangle$, $m\in \Z^d$, which satisfy
\be
\G |m_0,\dots,m_{d-1}\rangle= |m_0,\dots,-m_{d-1}\rangle.
\ee
In these notations, it is convenient to define a second orthonormal basis, where $m\in \Z^{d}$ but with $m_{d-1}\ge 1$:
\be
\begin{aligned}
&
|m_0,\dots,m_{d-2}, 0\rangle, \\
 &|m_0,\dots,m_{d-2}, m_{d-1}; \pm\rangle\equiv{1\over \sqrt{2}}\Big(|m_0,\dots,m_{d-2}, m_{d-1}\rangle\pm |m_0,\dots,m_{d-2}, - m_{d-1}\rangle\Big).
\end{aligned}
\ee
Because this new basis diagonalizes $\G$, 
\be
\G|m_0,\dots,m_{d-2}, 0\rangle=0,\quad \G|m_0,\dots,m_{d-2}, m_{d-1}; \pm\rangle=\pm |m_0,\dots,m_{d-2}, m_{d-1}; \pm\rangle,
\ee
we may translate Eq.~(\ref{pi2}) into
\be
\begin{aligned}
&\ln Z_{\rm B}=\ln Z_{\rm B}^{\rm e}+\ln Z_{\rm B}^{\rm o},\\
\where \quad &\ln Z_{\rm B}^{\rm e}=-{1\over 2}\, \mbox{Tr}\, {\boldsymbol{1}+\G\over 2} \ln\!\Big[\Big({m_\mu\over R_\mu}\Big)^2+M_{\rm B}^2\Big],\esp \\
&\ln Z_{\rm B}^{\rm o}=-{1\over 2}\, \mbox{Tr}\, {\boldsymbol{1}-\G\over 2} \ln\!\Big[\Big({m_\mu\over R_\mu}\Big)^2+M_{\rm B}^2\Big],
\end{aligned}
\label{rel}
\ee
which will turn out to be a suitable form for comparison with the string theory analysis to be presented in Sects~\ref{Orbi} and \ref{hb}.

\vskip.01in
\noindent
{\bf Statistical physics interpretation:} 
To present shortly a usual derivation of the Casimir effect, we can interpret the above results from a statistical physics viewpoint. We are interested in the large volume limit along the directions $x^1,\dots,x^{d-2}$, where the discrete modes become continuous. 
In this case, the path integral expression~(\ref{pi}) reads
\be
\begin{aligned}
&\ln Z_{\rm B}=-{1\over 2}\sum_{m_0}{V_{1,\dots,d-2}\over (2\pi)^{d-2}}\int d^{d-2}k\sum_{m_{d-1}}\ln\! \Big[\Big({m_0\over R_0}\Big)^2+\omega_{\mathbf{k}}^2\Big]\\
\where\quad & \omega_\mathbf{k}=\sqrt{\sum_{\mu=1}^{d-2}k_\mu^2+\Big({m_{d-1}\over R_{d-1}}\Big)^2+M_{\rm B}^2}\, .
\end{aligned}
\label{zth}
\ee
To make contact with thermodynamics,  we  write
\be
\sum_{m_0}\ln\! \Big[\Big({m_0\over R_0}\Big)^2+\omega_\mathbf{k}^2\Big]=2\ln\! \Big[2\sinh(\pi R_0\omega_\mathbf{k})\Big]+2\ln\! \Big[{1\over 2\pi R_0}\prod_{m_0\ge 1}\Big({m_0\over R_0}\Big)^2\Big],
\label{ze}
\ee
by employing the identity
\be
\sinh(x)= x\prod_{m_0\ge 1}\Big(1+{x^2\over \pi^2m_0^2}\Big).
\ee 
Using the $\zeta$-regularisation formulas
\be
\prod_{m_0\ge 1}{1\over R_0^2}={1\over R_0^{2\zeta(0)}}=R_0,\qquad \prod_{m_0\ge 1}m_0=e^{-\zeta'(0)}=\sqrt{2\pi},
\ee
the second logarithm  in the r.h.s. of  Eq.~(\ref{ze}) vanishes and we obtain 
\be
\begin{aligned}
-{\ln Z_{\rm B}\over 2\pi R_0}={V_{1,\dots, d-2}\over (2\pi)^{d-2}}\int d^{d-2}k&\sum_{m_{d-1}}{\omega_\mathbf{k}\over 2}\\&+{1\over 2\pi R_0}{V_{1,\dots, d-2}\over (2\pi)^{d-2}}\int d^{d-2}k\sum_{m_{d-1}}\ln\! \Big[1-e^{-2\pi R_0\omega_\mathbf{k}}\Big].
\end{aligned}
\label{free}
\ee
This result is nothing but the Helmholtz free energy of a perfect gas at finite temperature $T=1/(2\pi R_0)$. Note that the corresponding expression for a  fermionic degree of freedom could have been  obtained similarly if we had imposed an antiperiodic boundary condition along $S^1(R_0)$. Thus, the zero-temperature limit we are interested in is attained by taking  $R_0\to \infty$. In this limit, the second line vanishes and we are left with the vacuum energy $E$, whose density reads
\be
\V={E\over V_{1,\dots, d-1}}\equiv  -{\ln Z_{\rm B}\over V_{0,\dots, d-1}}={1\over 2\pi R_{d-1}}\int {d^{d-2}k\over (2\pi)^{d-2}}\sum_{m_{d-1}}{\omega_\mathbf{k}\over 2}.
\label{Vst}
\ee

For the orbifold cases, we may restrict to the cosine or sine modes and consider the length of the direction $x^{d-1}$ to be $\pi R_{d-1}$. Defining $V^{\rm h}_{1,\dots,d-1}=V_{1,\dots,d-1}/2$, the respective energy densities are
\be
\begin{aligned}
\V^{\rm e}={E^{\rm e}\over V^{\rm h}_{1,\dots, d-1}}\equiv  -{\ln Z_{\rm B}^{\rm e}\over V^{\rm half}_{0,\dots, d-1}}={1\over \pi R_{d-1}}\int {d^{d-2}k\over (2\pi)^{d-2}}\sum_{m_{d-1}\ge 0}{\omega_\mathbf{k}\over 2},\\
\V^{\rm o}={E^{\rm o}\over V^{\rm h}_{1,\dots, d-1}}\equiv  -{\ln Z_{\rm B}^{\rm o}\over V^{\rm h}_{0,\dots, d-1}}={1\over \pi R_{d-1}}\int {d^{d-2}k\over (2\pi)^{d-2}}\sum_{m_{d-1}\ge 1}{\omega_\mathbf{k}\over 2},
\end{aligned}
\label{veee1}
\ee
which obviously satisfy $2\V=\V^{\rm e}+\V^{\rm o}$.
Moreover, using the expression of $\ln Z_{\rm B}^{\rm e}$ in Eq.~(\ref{rel}), we have
\be
\V^{\rm e}=\V+\V_{\rm tw},\quad \where \quad \V_{\rm tw}={1\over 2}\,{\mbox{Tr}\, \G\ln \!\big[\big({m_\mu\over R_\mu}\big)^2+M_{\rm B}\big]\over V_{0,\dots,d-1}}.
\ee
However, evaluating the trace in the original orthonormal basis $|m\rangle$, $m\in\Z^d$, all contributions with $m_{d-1}\neq 0$ cancel out of $\V_{\rm tw}$, implying $\V_{\rm tw}$ to be  inversely proportional to $R_{d-1}$. In total, we obtain the final result
\be
\V^{\rm e}=\V+{\lambda_{\rm tw}\over R_{d-1}}, \quad \V^{\rm o}=\V-{\lambda_{\rm tw}\over R_{d-1}}, \quad \lambda_{\rm tw}={1\over 2}\,{\mbox{Tr}\, \ln \!\big[\sum_{\mu=0}^{d-2}\big({m_\mu\over R_\mu}\big)^2+M_{\rm B}\big]\over 2\pi\, V_{0,\dots,d-2}}.
\label{compa}
\ee


\paragraph{Alternative form:} 

The result~(\ref{Vst}) is clearly infinite if no cutoff is introduced. To extract the divergent part and simplify the computation of the Casimir force, we can present an alternative formulation of the energy density. Our starting point is Eq.~(\ref{zth}) where we exchange the roles of the Euclidean time $x^0_E$ and coordinate $x^{d-1}$,
\be
\begin{aligned}
&\ln Z_{\rm B}=-{1\over 2}\sum_{m_{d-1}}{V_{1,\dots,d-2}\over (2\pi)^{d-2}}\int d^{d-2}k\sum_{m_{0}}\ln\! \Big[\Big({m_{d-1}\over R_{d-1}}\Big)^2+\Omega_{k_{\rm E}}^2\Big]\\
\where\quad & \Omega_{k_{\rm E}}=\sqrt{\sum_{\mu=1}^{d-2}k_\mu^2+\Big({m_{0}\over R_{0}}\Big)^2+M_{\rm B}^2}\, .
\end{aligned}
\label{zth2}
\ee
Following identical steps, we  obtain the counterpart of Eq.~(\ref{free}),
\be
\begin{aligned}
-{\ln Z_{\rm B}\over 2\pi R_{d-1}}={V_{1,\dots, d-2}\over (2\pi)^{d-2}}\int d^{d-2}k&\sum_{m_{0}}{\Omega_{k_{\rm E}}\over 2}\\&+{1\over 2\pi R_{d-1}}{V_{1,\dots, d-2}\over (2\pi)^{d-2}}\int d^{d-2}k\sum_{m_{0}}\ln\! \Big[1-e^{-2\pi R_{d-1}\Omega_{k_{\rm E}}}\Big].
\end{aligned}
\ee
In the continuous limit \ie $R_0\to +\infty$, the density~(\ref{Vst}) takes the new form
\be
\begin{aligned}
\V\equiv & -{\ln Z_{\rm B}\over V_{0,\dots, d-1}}=\int {d^{d-1}k_{\rm E}\over (2\pi)^{d-1}} \,{\Omega_{k_{\rm E}}\over 2}+{1\over 2\pi R_{d-1}}\int {d^{d-1}k_{\rm E}\over (2\pi)^{d-1}}\ln\! \Big[1-e^{-2\pi R_{d-1}\Omega_{k_{\rm E}}}\Big],\quad \espD\\
&\where\quad  \Omega_{k_{\rm E}}=\sqrt{k_{\rm E}^2+M_{\rm B}^2},\quad k_{\rm E}\equiv (k_{{\rm E}0},\dots,k_{d-2}) .
\end{aligned}
\label{Vst2}
\ee
Notice that the second integral, which encodes the dependence on $R_{d-1}$, is convergent. Hence, the ill-defined part of $\V$ is the first integral, which is somehow an ``infinite constant''. As an example, we find for a vanishing mass, 
\be
M_{\rm B}=0: \quad \V=\int {d^{d-1}k_{\rm E}\over (2\pi)^{d-1}} \,{||k_{\rm E}||\over 2}-{v_d\over R_{d-1}^d}, \quad \where \quad v_d={\zeta(d)\over (2\pi)^{d}}\, {\Gamma\big({d\over 2}\big)\over \pi^{d\over 2}}.
\label{vd}
\ee


\paragraph{Casimir force:} 

Let us now consider the odd modes between the hyperplanes and compute the Casimir force in a usual way~\cite{IZ}. This can be done by considering the energy density between the plates when $R_{d-1}$ is finite, and when $R_{d-1}$ is very large (continuous limit). Using Eq.~(\ref{veee1}), they are respectively 
\be
\begin{aligned}
\V^{\rm o}&={1\over \pi R_{d-1}}\int {d^{d-2}k\over (2\pi)^{d-2}}\sum_{m_{d-1}\ge 1}{1\over 2}\, \sqrt{\sum_{\mu=1}^{d-2}k_\mu^2+\Big({m_{d-1} \over R_{d-1}}\Big)^2+M_{\rm B}^2},\\
\V^{\rm o}_\infty&={1\over \pi R_{d-1}}\int {d^{d-2}k\over (2\pi)^{d-2}}\; R_{d-1}\!\int_0^{+\infty} dk_{d-1}\, {1\over 2}\, \sqrt{\sum_{\mu=1}^{d-2}k_\mu^2+k_{d-1}^2+M_{\rm B}^2}.
\end{aligned}
\ee
Clearly, both are ill-defined due to the discrete sum and integrations over infinitely large momenta. However,  we may consider their difference, 
\be
\begin{aligned}
&\pi R_{d-1}\V^{\rm o}-(\pi R_{d-1}\V^{\rm o})_\infty={1\over 2}\int {d^{d-2}k\over (2\pi)^{d-2}}\\
&\times \left\{\sum_{m_{d-1}\ge 1}\sqrt{\sum_{\mu=1}^{d-2}k_\mu^2+\Big({m_{d-1} \over R_{d-1}}\Big)^2+M_{\rm B}^2}-R_{d-1}\!\int_0^{+\infty} dk_{d-1}\sqrt{\sum_{\mu=1}^{d-2}k_\mu^2+k_{d-1}^2+M_{\rm B}^2}\right\}\!,
\end{aligned}
\label{cres}
\ee
which can be evaluated by introducing a UV cutoff, and then sending the latter to infinity, which yields a finite result~\cite{IZ}. 

However, the final answer can be obtained more easily  by using our previous results. From Eq.~(\ref{compa}), we may write either  
\be
\V^{\rm o}-\V^{\rm o}_{\infty}=\V-{\lambda_{\rm tw}\over R_{d-1}}-\V_\infty \quad \mbox{or}\quad \pi R_{d-1}\V^{\rm o}-(\pi R_{d-1} \V^{\rm o})_{\infty}=\pi R_{d-1}\V-(\pi R_{d-1}\V)_\infty, 
\label{ov}
\ee
depending on which quantity, $\V^{\rm o}$ or $\pi R_{d-1}\V^{\rm o}$, we choose to take the $R_{d-1}\to +\infty$ limit. However, the second expression is more suitable, since $\lambda_{\rm tw}$ is actually infinite. Hence, we obtain for vanishing mass from Eq.~(\ref{vd}), 
\be
M_{\rm B}=0: \quad \pi R_{d-1}\V^{\rm o}-(\pi R_{d-1} \V^{\rm o})_{\infty}=-{\pi v_d\over R_{d-1}^{d-1}},
\label{fina}
\ee
which yields the force per unit area of the plates, \ie a pressure,
\be
\F\equiv-{\partial \big(R_{d-1}\V^{\rm o}-(R_{d-1}\V^{\rm o})_\infty\big)\over \partial R_{d-1}}=-(d-1)\, {v_d\over R_{d-1}^d}.
\label{fsdof}
\ee
This outcome is the Casimir force per unit area acting on each plate. In particular, it reproduces Eq.~(\ref{CFo}) after it is multiplied by 2 to account for the 2 degrees of freedom of the electromagnetic field in $d=4$, and for  $L=\pi R_{d-1}$. To understand why, let us note that on the hyperplane located at $x^{d-1}=\pi R_{d-1}$, there is a force arising from the vacuum comprised in the range $0<x^{d-1}<\pi R_{d-1}$, to which we must add a force induced by the vacuum in the infinite half space $x^{d-1}>\pi R_{d-1}$. Altogether, the resulting force per unit area acting on the plate is therefore 
\be
\F_{\rm int}+\F_{\rm ext}=-{\partial( R_{d-1}\V^{\rm o})\over \partial R_{d-1}} +{\partial (R_{d-1}\V^{\rm o}_\infty)\over \partial R_{d-1}}\equiv \F.
\label{fsdof2}
\ee

Hence, the Casimir force in field theory is the resultant of two formally infinite forces acting on each side of the plate, leaving a finite result. As we have just seen, this is equivalent to saying that the total force acting on a plate follows from the difference between the vacuum energies of two configurations, one in the presence of the plates at finite distance, and one with the second plate sent to infinity. This has the effect of removing the infinities encountered in the calculation of each configuration individually and is totally meaningful, as only differences in energies have a physical meaning when gravity is turned off. However, in presence of gravity, the situation is different, as the vacuum energy is nothing but the cosmological constant, and the latter cannot be infinite in a UV-complete theory.

%
%
\subsection{First quantized formalism with  spontaneous or explicit supersymmetry breaking} 
\label{ssBF}

To get closer to the string derivations to be given in the next sections, we may translate the above results in first quantized formalism. From Eq.~(\ref{pi}),  and inspired by the following equality
\be
-{1\over 2}\ln C = \int_0^{+\infty}{d\tau_2\over 2\tau_2}\, (e^{-\pi \tau_2C}-e^{-\pi\tau_2}),\quad \forall C>0,
\label{reg}
\ee
we may define the  regularized quantity
\be
\ln Z_{\rm B}^{\rm reg}=\int_\epsilon^{+\infty}{d\tau_2\over 2\tau_2}\sum_{m\in\Z^d} e^{-\pi\tau_2 \big[\big({m_\mu\over R_\mu}\big)^2+M_{\rm B}^2\big]}.  
\label{zreg}
\ee
In the above definition, the $C$-independent term of the integrand in Eq~(\ref{reg}) is not included, because empty of any information about the physical system. 
The dummy variable $\tau_2$ is the  Schwinger parameter, which is  the proper time of the particle along its trajectory, whose topology is that of a circle. In order to regulate the UV region $\tau_2\to 0$, a cutoff $\epsilon>0$ is introduced, and should sent to 0 only at the end of any sensible computation.  Notice that in this formulation, the discrete sum appearing in $\ln Z_{\rm B}^{\rm reg}$ can be taken over infinitely large momenta. Of course, the counterpart of $\epsilon$ in second quantized formalism, Eq.~(\ref{pi}), is a cutoff $\Lambda_{\rm co}^2>\sum_{\mu=0}^{d-1}(m_\mu/R_\mu)^2$.
As mentioned in the computation of the Casimir force in Eq.~(\ref{cres}), the latter should also be sent to infinity at the final step of a derivation. 
Hence, whether the first or second quantized formalism is used, only differences between two regulated energies associated with different configurations should be considered, in order to get finite (and equal) answers once the limits $\epsilon\to 0$ or $\Lambda_{\rm co}\to +\infty$ are taken.

However, compared to what we have discussed so far in field theory, the string theory case discussed  shortly will be different in two respects: It involves bosonic and fermionic degrees of freedom, and full towers of KK modes arising from a Scherk-Schwarz mechanism~\cite{SS,SS1,SSstring,s1,s2,s3,SSstring2}. For these reasons, we will not make use of the definition~(\ref{zreg}) and will follow another route, in our presentation of the first quantized formalism in quantum field theory. 


\paragraph{Scherk-Schwarz breaking of supersymmetry:} 

By considering the framework used in Sect.~\ref{secondQ} with one more circle denoted $S^1(R_9)$, and along which the quantum field has periodic or antiperiodic boundary conditions~\cite{SS,SS1}, 
\be
\forall x_E\in \R^d, \;\ \forall x^9\in \R,\quad \phi(x_E,x_9)=(-1)^\xi \phi(x_E,x^9+2\pi R_9), \quad \where \quad \xi=0 \mbox{ or } 1, 
\ee
the partition function Eq.~(\ref{pi}) in the massless case is generalized to 
\be
\ln Z_{\rm B}=-{1\over 2}\sum_{m\in \Z^d}\sum_{m_9}\ln\! \Big[\Big({m_\mu\over R_\mu}\Big)^2+\Big({m_9+{\xi\over 2}\over R_9}\Big)^2\Big].
\label{piKK}
\ee
In fact, we can also see the above formula as resulting from an infinite tower of KK modes with masses $M_{\rm B}={|m_9+{\xi\over 2} |\over R_9}$.  
To achieve the Scherk-Schwarz mechanism along $S^1(R_9)$, we can consider a massless fermionic free field with reversed boundary conditions along $S^1(R_9)$. The logarithm of the associated path integral is dressed with an overall minus sign,
\be
\ln Z_{\rm F}=+{1\over 2}\sum_{m\in \Z^d}\sum_{m_9}\ln\! \Big[\Big({m_\mu\over R_\mu}\Big)^2+\Big({m_9+{1-\xi\over 2}\over R_9}\Big)^2\Big].
\label{piKKf}
\ee
Note that the masses of the fermionic KK states are $M_{\rm F}={|m_9+{1-\xi\over 2} |\over R_9}$ so that the mass splitting between bosons and fermions at each level $m_9$ is identified to be 
\be
M={1\over 2R_9},
\label{msusy}
\ee
which can be interpreted as a scale of supersymmetry breaking. 

Using Eq.~(\ref{reg}) for the couple of such bosonic and fermionic fields, we obtain 
\be
\begin{aligned}
\ln Z_{\rm B+F} ={V_{0,\dots, d-2}\over (2\pi)^{d-1}}&\int d^{d-1}k_{\rm E}\sum_{m_{d-1}}\int_0^{+\infty} {d\tau_2\over 2\tau_2}\sum_{m_9}\\
&\bigg(e^{-\pi\tau_2\big[k_{\rm E}^2+\big({m_{d-1}\over R_{d-1}}\big)^2+\big({m_9+{\xi\over 2}\over R_9}\big)^2\big]}-e^{-\pi\tau_2\big[k_{\rm E}^2+\big({m_{d-1}\over R_{d-1}}\big)^2+\big({m_9+{1-\xi\over 2}\over R_9}\big)^2\big]}\bigg),
\end{aligned}
\label{aa}
\ee
where we have considered the continuous limit in the Euclidean directions $x^0_E,\dots,x^{d-2}$. Note that the $C$-independent term in the integrand of  Eq.~(\ref{reg}) drops by itself. Integrating over the continuous momenta, and applying Poisson resummations over the momenta $m_{d-1}$, $m_9$, we obtain
\be
\begin{aligned}
\ln Z_{\rm B+F} ={V_{0,\dots, d-2}\over (2\pi)^{d-1}}\, R_{d-1}R_9\sum_{\tilde m_{d-1}}&\int_0^{+\infty} {d\tau_2\over 2\tau_2^{1+{d+1\over 2}}}\sum_{\tilde m_9}\\
&e^{-{\pi \over \tau_2} \big[R_{d-1}^2\tilde m_{d-1}^2+R_9^2\tilde m_9^2\big]}(-1)^{\xi \tilde m_9}\Big[1-(-1)^{\tilde m_9}\Big],
\end{aligned}
\label{1q}
\ee
where the last bracket imposes $\tilde m_9$ to be odd, in order to contribute. As a result, the integral over $\tau_2$ is convergent, even in presence of the dangerous UV regime $\tau_2\to 0$, because of the exponential factor in the integrand. There is no need to introduce a cutoff of any kind, and the result can be written as
\be
\V_{\rm B+F}=-{\ln Z_{\rm B+F}\over V_{0,\dots,d-1}}=-(-1)^\xi f\big({\scriptstyle {R_{d-1}\over R_9}}\big) {1\over R_9^d},
\label{qft}
\ee 
where we have defined the function\footnote{Such functions have already  appeared in the literature in the expression of the effective potential at finite temperature derived in string theory models~\cite{crisol}. The latter describe cosmological evolutions which are attractors of the dynamics~\cite{attract,B1,B2,B3}. }
\be
f(u)={\Gamma\big({d+1\over 2}\big)\over (2\pi)^d\, \pi^{d+1\over 2}}\sum_{\tilde m_{d-1},\,  \tilde k_9} {1\over \big[(2\tilde k_9+1)^2+\tilde m_{d-1}^2u^2\big]^{d+1\over 2}}.
\label{f}
\ee
Of course the fact that we find a finite result without having to subtract anything to the energy is a consequence of the fact that we have considered the spontaneous breaking of a supersymmetric theory. In the decompactification limit $R_9\to +\infty$, supersymmetry is restored in $d+1$ dimensions. 

Notice that by  integrating over the continuous momenta, Eq.~(\ref{aa}) can be formally written  as
\be
\ln Z_{\rm B+F}=R_0\dots R_{d-2}\sum_{m_{d-1}}\int_0^{+\infty}{d\tau_2\over 2\tau_2^{1+{d-1\over 2}}}\, \Str e^{-\pi \tau_2 \big[\big({m_{d-1}\over R_{d-1}}\big)^2+{\cal M}^2\big]},
\ee
where the supertrace is over the bosonic and fermionic KK towers of states propagating along $S^1(R_9)$, and ${\cal M}$ is the associated KK mass operator. In this form, it is   straightforward to split the set of modes $m_{d-1}\in\Z$ relevant for a circular universe into those surviving  the orbifold projection $S^1(R_{d-1})/\Z_2$, and those involved in a Casimir-like universe, as explained below Eq.~(\ref{pi2}). In terms of projector and orthogonal projector, we obtain respectively 
\be
\begin{aligned}
\ln Z^{\rm e}_{\rm B+F}=R_0\dots R_{d-2}\sum_{m_{d-1}}\int_0^{+\infty}{d\tau_2\over 2\tau_2^{1+{d-1\over 2}}}\, \Str {\boldsymbol{1}+\G\over 2}\, e^{-\pi \tau_2 \big[\big({m_{d-1}\over R_{d-1}}\big)^2+{\cal M}^2\big]},\\
\ln Z^{\rm o}_{\rm B+F}=R_0\dots R_{d-2}\sum_{m_{d-1}}\int_0^{+\infty}{d\tau_2\over 2\tau_2^{1+{d-1\over 2}}}\, \Str {\boldsymbol{1}-\G\over 2}\, e^{-\pi \tau_2 \big[\big({m_{d-1}\over R_{d-1}}\big)^2+{\cal M}^2\big]},
\end{aligned}
\label{projs}
\ee
which are the analog of what will be computed in string theory. Proceeding as in the derivation of Eq.~(\ref{compa}) in second quantized formalism, we obtain
\be
\begin{aligned}
&\V^{\rm e}_{\rm B+F}=\V_{\rm B+F}+{\lambda_{\rm tw, B+F}\over R_{d-1}}, \quad\;\; \V^{\rm o}_{\rm B+F}=\V_{\rm B+F}-{\lambda_{\rm tw,B+F}\over R_{d-1}}, \\
\where \quad &\lambda_{\rm tw,B+F}={1\over (2\pi)^d}\int_0^{+\infty}{d\tau_2\over 2\tau_2^{1+{d-1\over 2}}}\, \Str e^{-\pi\tau_2{\cal M}^2}.
\end{aligned}
\label{compa2}
\ee
These quantities are all finite, and so are the forces per unit area $\F_{\rm int}$ and $\F_{\rm ext}$ exerted on either side of the plate located at $x^{d-1}=\pi R_{d-1}$,
\be
\F_{\rm int}= -{\partial (R_{d-1}\V^{\rm o}_{\rm B+F})\over \partial R_{d-1}}, \quad \;\; \F_{\rm ext}= -{\partial (R_{d-1}\V^{\rm o}_{\rm B+F})_\infty\over \partial R_{d-1}},
\ee
those resultant $\F=\F_{\rm int}+\F_{\rm ext}$ characterizes the Casimir effect. We will come back to its explicit  expression when we discuss the string theory point of view. 


\paragraph{Hard breaking of supersymmetry:} 

In order to construct a model realizing a ``hard breaking'' of supersymmetry, we are going to take the  limit $R_9\to 0$, which sends $M\to +\infty$. This is an arbitrary approach from the field theory point of view, which is efficient in the context of string theory, for further comparison. Starting from Eq.~(\ref{aa}), we integrate as before over the continuous momenta and Poisson resum over $m_{d-1}$. However, we Poisson resum over $m_9$ only when $\tilde m_{d-1}=0$, and keep the sum over $m_9$ as it is when $\tilde m_{d-1}\neq 0$.  As a result, the contribution $\tilde m_{d-1}=0$ is as in Eqs~(\ref{qft}),~(\ref{f}) and we obtain
\be
\begin{aligned}
\V_{\rm B+F}=   -(-1)^\xi\,  {\xi_{d}\over R_9^d} -{1\over (2\pi)^{d}}\sum_{\tilde m_{d-1}\neq 0}&\int_0^{+\infty} {d\tau_2\over 2\tau_2^{1+{d\over 2}}}\, e^{-{\pi \over \tau_2} R_{d-1}^2\tilde m_{d-1}^2}\\
&\sum_{m_9}\Big(e^{-\pi\tau_2\big({m_9+{\xi\over 2}\over R_9}\big)^2}-e^{-\pi\tau_2\big({m_9+{1-\xi\over 2}\over R_9}\big)^2}\Big),
\end{aligned}
\label{aaa}
\ee
where
\be
\xi_{d}={\Gamma\big({d+1\over 2}\big)\over (2\pi)^{d}\, \pi^{d+1\over 2}}\sum_{\tilde k_9}{1\over |2\tilde k_9+1|^{d+1}}.
\label{xid}
\ee
Clearly, the first term in Eq.~(\ref{aaa}), which is independent of $R_{d-1}$, is singular in the limit $R_9\to 0$. On the contrary, in the $R_{d-1}$-dependent part, the only contribution surviving the limit is that associated to the KK mode $m_9+{\xi\over 2}=0$ or  $m_9+{1-\xi\over 2}=0$, and it is finite. Hence, we may write, by abuse of notations,
\be
 \V_{\rm B+F}=-(-1)^\xi\,  {\xi_{d}\over R_9^d}-(-1)^\xi\, {v_d\over R_{d-1}^d}, \quad  \where \quad v_d={1\over 2(2\pi)^{d}}\, {\Gamma\big({d\over 2}\big)\over \pi^{d\over 2}}\sum_{\tilde m_{d-1}\neq 0}{1\over |\tilde m_{d-1}|^{d}},
\label{res1}
\ee
and where it is understood that $R_9$ is formally vanishing, making $\V_{\rm B+F}$ and $-(-1)^\xi \xi_{d}/R_9^d$ individually infinite. Alternatively, we can write both diverging quantities in the l.h.s. for their difference to be finite in the $R_9\to 0$ limit, 
\be
\V_{\rm B+F}-\V_{{\rm B+F},\infty}\equiv \V_{\rm B+F}+(-1)^\xi\,  {\xi_{d}\over R_9^d}=-(-1)^\xi\, {v_d\over R_{d-1}^d}.\label{res2}
\ee
Note that $v_d$ encodes the contributions of the KK excitations along $x^{d-1}$. Since the discrete sum appearing in  $v_d$ is $2\zeta(d)$, the result~(\ref{res1}) matches with Eq.~(\ref{vd}), where the interpretation of $v_d$ was obscure. 

In fact, in the limit $R_9\to 0$, all massive KK modes propagating along $S^1(R_9)$ decouple and we are left with a single massless boson (for $\xi=0$) or fermion (for $\xi=1$), with associated KK excitations along $x^{d-1}$, contributing to the energy density, however in an infinite way in $\V_{\rm B+F}$. There is no left notion of superpartner, thus realizing a hard, or explicit breaking of supersymmetry. Thus, the index ${\rm B+ F}$ is no more justified and should be replaced by either ${\rm B}$ of ${\rm F}$. Notice that the l.h.s. of Eq.~(\ref{res2}) amounts precisely to subtract the energy density at infinite $R_{d-1}$ to the result at finite $R_{d-1}$, to obtain a finite answer. Moreover, notice  that for a bosonic tower of KK modes, the ``infinite constant'' in Eq.~(\ref{vd}) is positive, while in Eq.~(\ref{res1}) it is negative ($\xi=0$). This is another occasion to stress that these equations are understood to make sense only when they are applied to evaluate the energy gap between two configurations. Under no circumstances, any of these equations should be used to evaluate some absolute energy, even in presence of a cutoff. As we will see in Sect.~\ref{hb}, this is where string theory is useful. 

Having discussed the case of a hard breaking of supersymmetry in a universe  circular along $x^{d-1}$, we can deduce the result once the  circular direction  is modded by $\Z_2$, and the modes are imposed to have nodes on the hyperplanes. From Eq.~(\ref{compa2}), we can evaluate $\lambda_{\rm tw,B+F}$ in the limit $R_9\to 0$, which yields  an infinite result (as in Eq.~(\ref{compa}) for $\lambda_{\rm tw}$),
\be
\lambda_{\rm tw,B+F}= {(-1)^\xi\over (2\pi)^d}\int_0^{+\infty}{d\tau_2\over 2\tau_2^{1+{d-1\over 2}}}.
\ee
Therefore, applying to $\V_{\rm B+F}^{\rm o}$ the second of the prescriptions given in Eq.~(\ref{ov}), we obtain 
\be
\pi R_{d-1}\V^{\rm o}_{\rm B+ F}-(\pi R_{d-1}\V^{\rm o}_{\rm B+ F})_\infty=-(-1)^\xi\,{\pi v_d\over R_{d-1}^{d-1}}, 
\ee
which yields the force per unit area 
\be
\F\equiv-{\partial \big(R_{d-1}\V^{\rm o}_{\rm B+ F}-(R_{d-1}\V^{\rm o}_{\rm B+ F})_{\infty}\big)\over \partial R_{d-1}}=-(-1)^\xi(d-1)\, {v_d\over R_{d-1}^d},
\label{fsdof3}
\ee
in  agreement with Eq~(\ref{fsdof}), for $\xi=0$


\section{Casimir effect  in string theory}
\label{Orbi}

The discussion of the Casimir effect in string theory we will present below, follows the logic we have developed in field theory, in first quantized formalism. In this section, we consider the case of a theory where supersymmetry is spontaneously broken \`a la Scherk-Schwarz, while the analysis where supersymmetry is explicitly broken will be presented  in Sect.~\ref{hb}.

\subsection{Orbifold realization and Scherk-Schwarz mechanism}

An appropriate framework for describing the case of a universe with one orbifolded direction $S^1(R_{d-1})/\Z_2$, where the modding acts as $x^{d-1}\to -x^{d-1}$, while {\em all degrees of freedom are imposed to the  even} under the transformation, is provided by the following background in heterotic string,
\be
\R^{1,d-2}\times {S^1(R_{d-1})\times T^3\over \Z_2}\times T^{6-d}\times S^1(R_9) .
\label{back}
\ee
In order for the vacuum energy not to be trivial, we implement a coordinate-dependent compactification along $S^1(R_9)$, which is nothing but a stringy version~\cite {SSstring,s1,s2,s3,SSstring2} of the Scherk-Schwarz mechanism~\cite {SS,SS1} responsible for the total spontaneous breaking of supersymmetry.  In string theory, the $\Z_2$ orbifold generator $\G$ acts as a twist on $S^1(R_{d-1})\times T^3$, namely 
\be
\G:\quad \left(X^{d-1},X^d,X^{d+1},X^{d+2}\right)\longrightarrow -\left(X^{d-1},X^d,X^{d+1},X^{d+2}\right)\!, 
\ee
where the $X$'s are the worldsheet bosonic fields realizing spatial directions. Unless specified, all formulas to come are valid for arbitrary radii $R_{d-1}$, $R_9$,  and arbitrary $T^3\times T^{6-d}$ torii moduli, provided 
\be
R_9>R_{\rm H}={\sqrt{2}+1\over \sqrt{2}}\quad \mbox{or}\quad R_9<{\sqrt{2}-1\over \sqrt{2}}={1\over 2R_H},
\label{hage}
\ee
for Hagedorn-like tachyonic instabilities not to occur. In practice, the directions of $T^3\times T^{6-d}\times S^1(R_9)$ should be seen as internal, while $\R^{1,d-2}\times S^1(R_{d-1})$ should be interpreted as spacetime.  However, due to the $\Z_2$-twist, the 10-dimensional background cannot be seen as a Cartesian product of internal and external spaces. The $(d-1)$-dimensional external space is  anisotropic, being restricted to sit between the  hyperplanes invariant under $\G$, \ie located at $X^{d-1}=0$ and $X^{d-1}=\pi R_{d-1}$.

Putting the non-compact directions of $\R^{1,d-2}$ in a ``box'', the regularized volume of spacetime is 
\be
V^{\rm half}_{0,\dots,d-1} = (2\pi R_0)\cdots (2\pi R_{d-2})\times \pi R_{d-1} .
\label{vol}
\ee
Following the  prescription as shown in the first line of Eq.~(\ref{projs}), the one-loop effective potential $\V^{\rm e}$ can be expressed in terms of the partition function in light-cone gauge, integrated over the fundamental domain $F$ of the modular group,
\begin{align}
\label{VV}
\!\!-V^{\rm half}_{0,\dots,d-1}\V^{\rm e}=&\!\int_F \!{d\tau_1d\tau_2\over2 \tau_2}\, {R_0R_1\over \tau_2}{R_2\cdots R_{d-2}\over(\sqrt{\tau_2}\eta\bar \eta)^{d-3}} \, {1\over 2}\sum_{H,G}Z_{(4,4)}\!\big[{}^H_G\big]  {\Gamma_{(6-d,6-d)}\over (\eta\bar \eta)^{6-d}}\, {R_9\over \sqrt{\tau_2}\eta\bar \eta}\sum_{n_9,\tilde m_9}e^{-{\pi R_9^2\over \tau_2}|\tilde m_9+n_9\tau|^2}\nonumber \\
&\,{1\over 2}\sum_{a,b}(-1)^{a+b+ab}{\theta[^a_b]^2\, \theta\big[{}^{a+H}_{b+G}\big]\, \theta\big[{}^{a-H}_{b-G}\big]\over \eta^4}\, {1\over 2}\sum_{\gamma,\delta}{\bar \theta[^\gamma_\delta]^6\, \bar \theta\big[{}^{\gamma+H}_{\delta+G}\big]\, \bar \theta\big[{}^{\gamma-H}_{\delta-G}\big]\over \bar \eta^8}\, {1\over 2}\sum_{\gamma',\delta'}{\bar \theta\big[{}^{\gamma'}_{\delta'}\big]^8\over \bar \eta^8}\nonumber \\
&\, (-1)^{\tilde m_9a+n_9b+\tilde m_9n_9}\, (-1)^{\xi(\tilde m_9\gamma+n_9\delta+\tilde m_9n_9)}\, (-1)^{\xi'(\tilde m_9\gamma'+n_9\delta'+\tilde m_9n_9)}.
\end{align}
Our notations are as follows~\cite{kiritsis}:

$\bullet$ $\tau=\tau_1+i\tau_2$ is the Teichm\"uller parameter of the genus-1 Riemann surface, while $\eta(\tau)$ and $\theta [^\alpha_\beta](\tau)$ are the Dedekind and Jacobi modular forms.

$\bullet$ The spin structures $a,b$ on the worldsheet, as well as the indices $\gamma,\delta$ and $\gamma',\delta'$ take values 0 or 1. 

$\bullet$ The lattice of zero modes associated to $S^1(R_9)$ is presented in  Lagrangian form, with $n_9, \tilde m_9\in\Z$, while that for a torus $T^n$ is denoted $\Gamma_{(n,n)}$.

$\bullet$ The conformal block associated to the twisted coordinates of $(S^1(R_{d-1})\times T^3)/\Z_2$ is
\be
\label{z44}
Z_{4,4}\!\big[{}^H_G\big]\!= \left\{
\begin{array}{ll}
\displaystyle {R_{d-1}\over \sqrt{\tau_2}}\sum_{n_{d-1},\tilde m_{d-1}}\!\!\!e^{-{\pi R_{d-1}^2\over \tau_2}|\tilde m_{d-1}+n_{d-1}\tau|}\, {\Gamma_{3,3}\over \eta^4\bar\eta^4}&\quad \mbox{if $(H,G)=(0,0)$} , \espDD\\
\displaystyle {2^4\, \eta^2\bar\eta^2 \over \theta\big[{}^{1-H}_{1-G}\big]^2\, \bar\theta\big[{}^{1-H}_{1-G}\big]^2}&\quad \mbox{if $(H,G)\neq (0,0)$}, \esp
\end{array}
\right.
\ee 
where $H,G$ take values 0 or 1. In particular, string theory contains a twisted sector $H=1$ which has no counterpart in field theory. In the sector $(H,G)=(0,0)$, the lattice of zero modes associated to $S^1(R_{d-1})$ is presented in Lagrangian form, with $n_{d-1},\tilde m_{d-1}\in \Z$.

$\bullet$ In the last line of Eq.~(\ref{VV}), the first sign depends on the spin structures $a,b$ and induces the super-Higgs mechanism~\cite{SSstring2}. The mass $M$ of the gravitini is as given in Eq.~(\ref{msusy}), which is  the scale 
of spontaneous supersymmetry breaking. 
In the second and third signs, we have introduced discrete parameters $\xi$ and $\xi'$ equal to 0 or 1, which implement a Higgs mechanism of the $E_8\times E_8$ gauge symmetry when they are not both vanishing. Choosing $\xi=1$ ($\xi'=1$) enforces the  $E_8\to SO(16)$ spontaneous breaking of the first (second) $E_8$ gauge group factor. Together, super-Higgs and Higgs mechanisms combine to  yield 4 different patterns of breakings: For $\xi=\xi'=0$, all initially massless fermions of the parent supersymmetric model (obtained when  the sign $(-1)^{\tilde m_9a+n_9b+\tilde m_9n_9}$ is omitted) acquire  a mass $1/(2R_9)$, while their bosonic superpartners remain massless. However, a non-trivial $\xi$ and/or $\xi'$ results in reversing the roles of bosons and fermions in supermultiplets, thus giving a mass to some bosons while maintaining  their fermionic partners massless~\cite{crisol,SNSM,SNSM2}. 

By noticing that in $\V^{\rm e}$, the sectors $(H,G)\neq (0,0)$ are inversely proportional to $R_{d-1}$, we may split the potential energy density as 
\be
\V^{\rm e}=\V(R_{d-1})+{\lambda_{\rm tw}\over R_{d-1}}.
\label{veee}
\ee
The first contribution,
\begin{align}
\V=&-{R_9\over 2(2\pi)^d}\int_F {d\tau_1d\tau_2\over \tau_2^{1+{d+1\over 2}}} \, \sum_{n_{d-1},\tilde m_{d-1}}\!\!\!e^{-{\pi R_{d-1}^2\over \tau_2}|\tilde m_{d-1}+n_{d-1}\tau|}\nonumber \\
&\,  \sum_{n_9,\tilde m_9}e^{-{\pi R_9^2\over \tau_2}|\tilde m_9+n_9\tau|^2}\, {1\over 2}\sum_{a,b}(-1)^{a+b+ab}\theta[{}^a_b]^4(-1)^{\tilde m_9a+n_9b+\tilde m_9n_9}\label{VV2} \\
& \, {1\over 2}\sum_{\gamma,\delta}\bar \theta\big[{}^\gamma_\delta\big]^8(-1)^{\xi(\tilde m_9\gamma+n_9\delta+\tilde m_9n_9)}\, {1\over 2}\sum_{\gamma',\delta'}\bar \theta\big[{}^{\gamma'}_{\delta'}\big]^8(-1)^{\xi'(\tilde m_9\gamma'+n_9\delta'+\tilde m_9n_9)}\, {\Gamma_{(3,3)}\Gamma_{(6-d,6-d)}\over \eta^{12}\bar\eta^{24}},\nonumber 
\end{align}
is actually the effective potential  arising in the model when no orbifold twist is implemented. Its expression is that found in the $d$-dimensional background periodic along $x^{d-1}$,
\be
\R^{1,d-2}\times S^1(R_{d-1})\times T^3\times T^{6-d}\times S^1(R_9),
\ee
with Scherk-Schwarz mechanism implemented along $S^1(R_9)$. The second term in Eq.~(\ref{veee}) is given by 
\be
\begin{aligned}
\!\!\lambda_{\rm tw}=&-{R_9\over 2(2\pi)^d}\int_F {d\tau_1d\tau_2\over \tau_2^{1+{d\over 2}}} \, {1\over 2}\sum_{a,b}(-1)^{a+b+ab}\sum_{n_9,\tilde m_9}e^{-{\pi R_9^2\over \tau_2}|\tilde m_9+n_9\tau|^2}(-1)^{\tilde m_9a+n_9b+\tilde m_9n_9} \\
& \,{1\over 2}\sum_{\gamma,\delta}(-1)^{\xi(\tilde m_9\gamma+n_9\delta+\tilde m_9n_9)}\, {1\over 2}\sum_{\gamma',\delta'}(-1)^{\xi'(\tilde m_9\gamma'+n_9\delta'+\tilde m_9n_9)}\, \Gamma_{(6-d,6-d)}\!\!\!\!\!\sum_{(H,G)\neq (0,0)}\!\!\!\!\!\omega\big[{}^{H,a,\gamma,\gamma'}_{G,\, b,\, \delta,\, \delta'}\big],
\end{aligned}
\label{aaaa}
\ee
where we have defined 
\be
\omega\big[{}^{H,a,\gamma,\gamma'}_{G,\, b,\, \delta,\, \delta'}\big]={1\over \eta^8\bar \eta^{20}}\, {2^4\, \eta^2\bar\eta^2 \over \theta\big[{}^{1-H}_{1-G}\big]^2\, \bar\theta\big[{}^{1-H}_{1-G}\big]^2} \, \theta[^a_b]^2\, \theta\big[{}^{a+H}_{b+G}\big]\, \theta\big[{}^{a-H}_{b-G}\big]\, \bar \theta[^\gamma_\delta]^6\,\bar \theta\big[{}^{\gamma+H}_{\delta+G}\big]\, \bar \theta\big[{}^{\gamma-H}_{\delta-G}\big]\, \bar \theta\big[{}^{\gamma'}_{\delta'}\big]^8.
\ee

To describe the case of a universe with one orbifolded direction $S^1(R_{d-1})/\Z_2$, while {\em all degrees of freedom are imposed to the odd} under the $\Z_2$ transformation, we may consider the  string theory orbifold model based on the {\em orthogonal} projector, as compared to the above described case,
\be
{\boldsymbol{1}-\G\over 2}=\boldsymbol{1}-{\boldsymbol{1}+\G\over 2},
\label{orbifolds}
\ee
exactly as developed in the quantum field theory framework, Eq.~(\ref{projs}). To write the associated effective potential $\V^{\rm o}$, notice that the flipped sign in front of $\G$ implies the sector $(H,G)=(0,1)$ to appear with an opposite sign, as compared to the case discussed so far. Given that the sectors $(H,G)\neq (0,0)$ realize a modular orbit of $SL(2,\Z)$, we conclude that all of them show up with an opposite sign. As a result, we obtain 
\be
\V^{\rm o}=\V(R_{d-1})-{\lambda_{\rm tw}\over R_{d-1}},
\ee
which can also be seen as a direct consequence of the equality in Eq.~(\ref{orbifolds}).

The remaining contribution $(H,G)= (0,0)$ being  modular invariant by itself, we may ``unfold'' in the expression of $\V$ the domain of integration $F$ into the ``upper-half strip'', by using the following result \cite{McClain:1986id,M1}:   For any set of  functions $v_{(n,\tilde m)}(\tau,\bar \tau)$ satisfying 
\be
\forall M=\begin{pmatrix}a & b \\ c & d \end{pmatrix} \in SL(2,\Z),\;\; v_{(n,\tilde m)}\Big({a\tau+b\over c\tau+d},{a\bar \tau+b\over c\bar \tau+d}\Big)=v_{(n,\tilde m)M}(\tau,\bar \tau),  
\ee
the sum $\sum_{n,\tilde m}v_{(n,\tilde m)}$ is modular invariant and, provided the discrete series are  absolutely convergent, for the exchange of  the discrete sum and integration to be legitimate,  we have 
\be
\label{unfo}
\int_F{d\tau_1d\tau_2\over \tau_2^2}\, \sum_{n,\tilde m}v_{(n,\tilde m)}(\tau,\bar \tau)=\int_F{d^2\tau\over \tau_2^2}\, v_{(0,0)}(\tau,\bar \tau)+\int_{-{1\over 2}}^{1\over 2}d\tau_1\int_0^{+\infty}{d\tau_2\over \tau_2^2}\, \sum_{\tilde m\neq 0}v_{(0,\tilde m)}(\tau,\bar \tau).
\ee
In our case at hand, this equality turns out to be valid for the series of functions labeled by $(n_{d-1},\tilde m_{d-1})$ when $R_{d-1}>1$, which we know is satisfied since $x^{d-1}$ is an external space direction. As result, we obtain
\be
\V^{\rm o}=\V_0+\V_*(R_{d-1})-{\lambda_{\rm tw}\over R_{d-1}},
\label{Vfull}
\ee
where the  first contribution is independent of $R_{d-1}$,
\be
\label{LR9}
\begin{aligned}
\!\!\!\V_0 =& -{R_9\over 2(2\pi)^d}\int_F {d\tau_1d\tau_2\over \tau_2^{1+{d+1\over 2}}} \, \sum_{n_9,\tilde m_9}e^{-{\pi R_9^2\over \tau_2}|\tilde m_9+n_9\tau|^2}\, {1\over 2}\sum_{a,b}(-1)^{a+b+ab}\theta[{}^a_b]^4(-1)^{\tilde m_9a+n_9b+\tilde m_9n_9} \\
& \,  {1\over 2}\sum_{\gamma,\delta}\bar \theta\big[{}^\gamma_\delta\big]^8(-1)^{\xi(\tilde m_9\gamma+n_9\delta+\tilde m_9n_9)}\, {1\over 2}\sum_{\gamma',\delta'}\bar \theta\big[{}^{\gamma'}_{\delta'}\big]^8(-1)^{\xi'(\tilde m_9\gamma'+n_9\delta'+\tilde m_9n_9)}\, {\Gamma_{(3,3)}\Gamma_{(6-d,6-d)}\over \eta^{12}\bar\eta^{24}},
\end{aligned}
\ee
whereas the second depends non-trivially on $R_{d-1}$, 
\begin{align}
\V_* =& -{R_9\over 2(2\pi)^d}\int_{-{1\over 2}}^{1\over 2}d\tau_1\int_0^{+\infty} {d\tau_2\over \tau_2^{1+{d+1\over 2}}} \, \sum_{\tilde m_{d-1}\neq 0}e^{-{\pi R_{d-1}^2\over \tau_2}\tilde m_{d-1}^2}\nonumber \\
&\,  \sum_{n_9,\tilde m_9}e^{-{\pi R_9^2\over \tau_2}|\tilde m_9+n_9\tau|^2}\, {1\over 2}\sum_{a,b}(-1)^{a+b+ab}\theta[{}^a_b]^4(-1)^{\tilde m_9a+n_9b+\tilde m_9n_9}\label{VQFT} \\
& \, {1\over 2}\sum_{\gamma,\delta}\bar \theta\big[{}^\gamma_\delta\big]^8(-1)^{\xi(\tilde m_9\gamma+n_9\delta+\tilde m_9n_9)}\, {1\over 2}\sum_{\gamma',\delta'}\bar \theta\big[{}^{\gamma'}_{\delta'}\big]^8(-1)^{\xi'(\tilde m_9\gamma'+n_9\delta'+\tilde m_9n_9)}\, {\Gamma_{(3,3)}\Gamma_{(6-d,6-d)}\over \eta^{12}\bar\eta^{24}}.\nonumber 
\end{align}
Due to the integration over $\tau_1$, $\V_*$  involves only the level-matched spectrum of the string. Notice that it is also finite, due to the presence of the factor $\exp(-\pi R_{d-1}^2\tilde m_{d-1}^2/\tau_2)$, which allows the integration even in the dangerous UV region $\tau_2\to 0$. 

As a result, the force per unit area arising from the vacuum  comprised in the range $0<x^{d-1}<\pi R_{d-1}$ and acting on the boundary hyperplane located  at $x^{d-1}=\pi R_{d-1}$ is
\be
\F_{\rm int}=-{\partial (R_{d-1}\V^{\rm o})\over \partial R_{d-1}}=-\V_0-{\partial (R_{d-1}\V_*)\over \partial R_{d-1}}.
\label{finter}
\ee
This is the end of the story if we want to truly interpret the hyperplanes as the boundaries of the universe. However, applying the result to describe the Casimir effect, we need to add the force per unit area arising from the infinite half space $x^{d-1}>\pi R_{d-1}$. Up to a sign, it is equal to $\F_{\rm int}$ evaluated in the limit $R_{d-1}\to +\infty$, as the second plate is sent to infinity. Since $\V_*$ vanishes in this limit, we have 
\be
\F_{\rm ext}=\V_0,
\label{fexter}
\ee
and the Casimir force per unit area is
\be
\F\equiv \F_{\rm int}+\F_{\rm ext}=-{\partial (R_{d-1}\V_*)\over \partial R_{d-1}}.
\label{casist}
\ee
It is independent of the twisted sector, $H=1$, of the string spectrum. This is not a surprise since the latter is restricted to live on the hyperplanes fixed by the orbifold projection. Having no modes propagating in the transverse direction $x^{d-1}$, they cannot contribute to the force.

Comparing the result with that found for a single   degree of freedom in field theory, Eqs~(\ref{fsdof2}) or~(\ref{fsdof3}), we see that in the string theory framework, the Casimir force is not derived by subtracting two infinite quantities to end up with a finite result. Actually, both $\V_*$ and $\V_0$ are finite, where $\V_0$ is the analogue of what was denoted $\V^{\rm o}_\infty$ (or $\V^{\rm o}_{\rm B+F,\infty}$) in field theory. However, one may think that the finiteness of $\V_0$ arises only because we considered a string theory model where supersymmetry is spontaneously broken \`a la Scherk-Schwarz, since we have shown in Eq.~(\ref{qft}) that finiteness is also encountered in field theory for each pair of non-degenerate superpartners with gap $M$ between their masses. To clarify this issue, we now compare in details the two frameworks when they both realize a spontaneous breaking of supersymmetry, and will show in Sect.~\ref{hb} that finiteness survives in string theory {\em even when supersymmetry is explicitly broken}. 

\subsection{String theory {\em versus} quantum field theory}

Our aim is to figure out the differences between the finite field theory result $\V_{\rm B+F}$ given in Eq.~(\ref{qft}), and that found in string theory, $\V_0+\V_*$, which are both associated with configurations where supersymmetry breaking is spontaneously broken \via a Scherk-Schwarz mechanism. To be specific, for the notion of spontaneous breaking to be valid in the string theory model, let us assume in this subsection $R_9>R_H$ in Eq.~(\ref{hage}), in order for  $R_9$ to be allowed to take very large values and possibly restore supersymmetry (in $d+1$ dimensions). 

Let us first consider  $\V_0$ on which we can apply the unfolding formula, Eq.~(\ref{unfo}),  for the series of functions labeled by $(n_9,\tilde m_9)$. This can be done since absolute convergence is valid for $R_9>R_{\rm H}$. The contribution for $(n_9,\tilde m_9)=(0,0)$, which is integrated over the fundamental domain $F$, vanishes due to the exact supersymmetry of the integrand. For the same reason, the contributions $(n_9,\tilde m_9)=(0,2\tilde k_9)$, $\tilde k_9\in\Z^*$, which are integrated over the upper-half strip, vanish as well. Thus, we obtain 
\be
\begin{aligned}
\V_0 =& -{R_9\over 2(2\pi)^d}\int_{-{1\over 2}}^{1\over 2}d\tau_1\int_0^{+\infty} {d\tau_2\over \tau_2^{1+{d+1\over 2}}} \, \sum_{\tilde k_9}e^{-{\pi R_9^2\over \tau_2}(2\tilde k_9+1)^2}\, {1\over 2}\sum_{a,b}(-1)^{a+b+ab}\theta[{}^a_b]^4(-1)^{a} \\
& \, {1\over 2}\sum_{\gamma,\delta}\bar \theta\big[{}^\gamma_\delta\big]^8(-1)^{\xi\gamma}\, {1\over 2}\sum_{\gamma',\delta'}\bar \theta\big[{}^{\gamma'}_{\delta'}\big]^8(-1)^{\xi'\gamma'}\, {\Gamma_{(3,3)}\Gamma_{(6-d,6-d)}\over \eta^{12}\bar\eta^{24}},
\label{Lambda}
\end{aligned}
\ee
which reduces to an expression involving the level-matched spectrum only, due to the integration over $\tau_1$.

The field theory we have considered in Sect.~\ref{ssBF} involves one massless boson (fermion) and one fermion (boson) of mass $M$, together with their KK towers of states propagating along the internal circle $S^1(R_9)$. However, beside such KK towers, the physical spectrum contributing to $\V_0$ also contains states that cannot be organized as pure KK towers of modes associated to  $S^1(R_9)$. The latter involve momentum and/or winding modes along all internal directions and/or from string oscillators.  To make contact with the field theory result, we thus restrict from now on to the case where $M$ is much lower than all other mass scales present in the model.  In particular, we have $R_9\gg 1$, for the string scale to be much heavier than $M$. In practice, the lightest states of the model are therefore nothing but the KK modes propagating along $S^1(R_9)$ and they will dominate in the expression of $\V_0$. To see this, we expand 
\be
\label{expand}
\begin{aligned}
&{1\over 2}\sum_{a,b}(-1)^{a+b+ab}{\theta[{}^a_b]^4\over \eta^{12}}(-1)^{a} = {\theta\!\big[{}^1_0\big]^4\over \eta^{12}}=16\big(1+\O(q)\big), \\
 &{1\over 2}\sum_{\gamma,\delta}\bar \theta\big[{}^\gamma_\delta\big]^8(-1)^{\xi\gamma}=1+112\bar q+(-1)^\xi 128 \bar q+\O(\bar q^2),\\
& {1\over \bar \eta^{24}}={1\over \bar q}\big(1+24\bar q+\O(\bar q^2)\big), \espD\\
& \Gamma_{(3,3)}\Gamma_{(6-d,6-d)}=1+d_{\rm en}\bar q +\O(e^{-\pi c^2\tau_2}), 
\end{aligned}
\ee
where $q=e^{2i\pi\tau}$.
In the last line, we take into account the possibility that the internal metric and antisymmetric tensor appearing in the $\Gamma_{(3,3)}$ and $\Gamma_{(6-d,6-d)}$ lattices sit at an enhanced symmetry point in moduli space, thus yielding $d_{\rm en}$ additional massless vector multiplets in the supersymmetric parent theory.  Moreover, a moduli-dependent constant $c$ is introduced to account for the fact that the mass scales introduced by the lattices are heavier than the supersymmetry breaking scale, \ie that $1>c\gg 1/R_9$. By using the following result, valid for $K,{\cal M}>0$,
\be
 \int_0^{+\infty}{d\tau_2\over \tau_2^{1+{d+1\over 2}}}\, e^{-{\pi\over \tau_2}K^2}\, e^{-\pi \tau_2 {\cal M}^2}\sim {{\cal M}^{d\over 2}\over K^{1+{d\over 2}}}\, e^{-2\pi K{\cal M}}\quad \when\quad  K{\cal M}\gg1,
 \label{supp}
\ee
we obtain
\be
\V_0= (\nF-\nB)\,  {\xi_{d}\over R_9^d}+\O\!\left(\!\Big({c\over R_9}\Big)^{d\over 2}e^{-2\pi cR_9}\right)\!,
\label{v0fin}
\ee
where $\xi_d$ is defined in Eq.~(\ref{xid}). In the above expression, $\nF$ and $\nB$ are the numbers massless fermionic and  bosonic degrees of freedom, and they satisfy
\be
\nF-\nB= 8\big((-1)^{\xi+1} 128+(-1)^{\xi'+1}128-248-d_{\rm en}\big) .
\label{vonfb}
\ee
Notice that the dressing coefficient $\xi_d$ in Eq.~(\ref{v0fin}) reflects  the fact that their entire KK towers of states associated to the large circle $S^1(R_9)$ contribute significantly. On the contrary, all other  string states yield  exponentially suppressed contributions to $\V_0$. 
Note that the sign of $\V_0$ can be positive or negative: $(\xi,\xi')\neq (1,1)$ implies $\nF-\nB<0$, while $(\xi,\xi')= (1,1)$, $d_{\rm en}\le 8$ yields  $\nF-\nB\ge 0$. 

The expression of $\V_*$ in Eq.~(\ref{VQFT}) can also be simplified.  Applying a Poisson resummation over $\tilde m_9$, the lattice of zero modes associated to the Scherk-Schwarz circle can be written in Hamiltonian form,
\begin{align}
&{R_9\over \sqrt{\tau_2}}\sum_{\tilde m_9}e^{-{\pi R_9^2\over \tau_2}|\tilde m_9+n_9\tau|^2}(-1)^{\tilde m_9[a+\xi\gamma+\xi'\gamma'+n_9(1+\xi\gamma+\xi'\gamma')]} = \sum_{m_9}q^{{1\over 4}\left({M_9\over R_9}+n_9 R_9\right)^2}\bar q^{{1\over 4}\left({M_9\over R_9}-n_9 R_9\right)^2}, \nonumber  \\
&\where \qquad M_9=m_9+{1\over 2}\big[a+\xi\gamma+\xi'\gamma'+n_9(1+\xi+\xi ')\big].
\label{H9}
\end{align}
Applying  Eq.~(\ref{supp}) for $K=R_{d-1}|\tilde m_{d-1}|$ and any mass ${\cal M}= \sqrt{(n_9R_9)^2+\cdots}$, where the ellipses are positive, one concludes that since $R_9\gg 1$, the contributions for $n_9\neq 0$ are exponentially suppressed, as compared to those arising for $n_9=0$. In that case, we may write 
\begin{align}
\V_* =& -{R_9\over 2(2\pi)^d}\int_{-{1\over 2}}^{1\over 2}d\tau_1\int_0^{+\infty} {d\tau_2\over \tau_2^{1+{d+1\over 2}}} \, \sum_{\tilde m_{d-1}\neq 0}e^{-{\pi R_{d-1}^2\over \tau_2}\tilde m_{d-1}^2}\, \sum_{\tilde k_9}e^{-{\pi R_9^2\over \tau_2}(2\tilde k_9+1)^2}\nonumber \\ 
&\, {1\over 2}\sum_{a,b}(-1)^{a+b+ab}\theta[{}^a_b]^4(-1)^{a} \, {1\over 2}\sum_{\gamma,\delta}\bar \theta\big[{}^\gamma_\delta\big]^8(-1)^{\xi\gamma}\, {1\over 2}\sum_{\gamma',\delta'}\bar \theta\big[{}^{\gamma'}_{\delta'}\big]^8(-1)^{\xi'\gamma'}\,   {\Gamma_{(3,3)}\Gamma_{(6-d,6-d)}\over \eta^{12}\bar\eta^{24}}\nonumber \\
&\, +\O\left(e^{-2\pi R_{d-1}R_9}\right),
\end{align}
where we have used the fact that only odd values of $\tilde m_9=2\tilde k_9+1$ contribute, due to supersymmetry.  In fact, $\V_0$ given in Eq.~(\ref{Lambda}) is nothing but the ``missing'' $\tilde m_{d-1}=0$ term of the above dominant contribution of $\V_*$. By proceeding exactly as we did for evaluating  $\V_0$, we find
\be
\V_*= (\nF-\nB)\, f_*\big({\scriptstyle {R_{d-1}\over R_9}}\big)\,  {1\over R_9^d}+\O\!\left(\!\Big({c\over R_9}\Big)^{d\over 2}e^{-2\pi c\sqrt{R_9^2+R_{d-1}^2}}\right)\!,
\label{v*fin}
\ee
where we have defined
\be
f_*(u)={\Gamma\big({d+1\over 2}\big)\over 2^{d}\, \pi^{3d+1\over 2}}\sum_{\tilde m_{d-1}\neq 0,\,  \tilde k_9} {1\over \big[(2\tilde k_9+1)^2+\tilde m_{d-1}^2u^2\big]^{d+1\over 2}}.
\ee
Combining Eqs~(\ref{v0fin}) and~(\ref{v*fin}), the final expression in string theory  of the energy density in a universe circular  along $x^{d-1}$ turns out to be in  agreement with the field theory analysis, when supersymmetry is spontaneously broken by the Scherk-Schwarz mechanism at a low scale. The finite result is,\footnote{Such expressions also appear in  the context of cosmological solutions~\cite{crisol,cr1,attract} or in models with vanishing effective potential at one-loop~\cite{SNSM,SNSM2}.}  
\be
\V_0+\V_*=(\nF-\nB)\, f\big({\scriptstyle {R_{d-1}\over R_9}}\big)\,  {1\over R_9^d}+\O\!\left(\!\Big({c\over R_9}\Big)^{d\over 2}e^{-2\pi c\sqrt{R_9^2+R_{d-1}^2}}\right)\!,
\ee
which matches with Eq.~(\ref{qft}) for $\nB$ and $\nF$ massless bosonic and fermionic degrees of freedom and  their KK towers of modes, up to exponentially suppressed contributions arising from heavy string states not present in field theory. Hence, the Casimir force can be derived in both frameworks from Eqs.~(\ref{casist}) and~(\ref{v*fin}).


\section{Casimir effect and hard breaking of supersymmetry}
\label{hb}

So far, we have analyzed the Casimir effect in the case of a spontaneous breaking of supersymmetry. We observed that in string and field theory, the force acting on the hyperplane located at $x^{d-1}=0$ and generated by the vacuum energy between the two plates is finite, and so is the force induced by the vacuum in the infinite space $x^{d-1}<0$. In the limit of low supersymmetry breaking scale, as compared to the other scales arising in the string model (Higgs-like or string scale), the two frameworks yield identical results. Therefore, string theory is of limited interest in this case, since the field theory approach is technically  simpler.  However, the situation may be drastically different in models where supersymmetry is explicitly broken, due to the illness of the definition of the energy density in field theory in these conditions. As seen in Eq.~(\ref{ov}) (or in Eqs~(\ref{fina}) and~(\ref{res2}) for the massless case) a prescription consisting in subtracting the energies associated to two configurations, though individually divergent, is required to make sense.  

To analyse the string theory picture when supersymmetry breaking is hard, we note that Eq.~(\ref{Vfull}) is valid for both ranges defined in Eq.~(\ref{hage}), which are related by T-duality $R_9\to 1/(2R_9)$. Thus, instead of considering the model of the previous section at large $R_9$ (and even recovering exact supersymmetry in $d+1$ dimensions as $R_9\to +\infty$), we may consider the opposite limit, $R_9\to 0$, where the scale of supersymmetry breaking and actually the mass of the gravitinos are sent to infinity. The gravitini being decoupled from the rest of the spectrum, explicit and consistent models in $d+1$ spacetime dimensions  with hard breaking of supersymmetry may be found this way, provided no tachyons are generated in taking the limit~\cite{o16}. 

In the expressions of $\V_0$, $\V_*$ and $\lambda_{\rm tw}$ given in Eqs~(\ref{LR9}),~(\ref{VQFT}) and~(\ref{aaaa}), let us redefine $\tilde m_9\equiv 2\tilde k_9+g$ and $n_9\equiv 2l_9+h$, where $g,h\in\{0,1\}$, and perform Poisson resummations over $\tilde k_9$ and $l_9$. Including the signs responsible for super-Higgs and the Higgs mechanisms, the relevant terms are
\begin{align}
&R_9\sum_{n_9,\tilde m_9}e^{-{\pi R_9^2\over \tau_2}|\tilde m_9+n_9\tau|^2}(-1)^{\tilde m_9a+n_9b+\tilde m_9n_9}\, (-1)^{\xi(\tilde m_9\gamma+n_9\delta+\tilde m_9n_9)}\, (-1)^{\xi'(\tilde m_9\gamma'+n_9\delta'+\tilde m_9n_9)}\nonumber \\
&={1\over 2}\sum_{h,g}2R_9\sum_{l_9,\tilde k_9}e^{-{\pi (2R_9)^2\over \tau_2}|\tilde k_9+{g\over 2}+(l_9+{h\over 2})\tau|^2}(-1)^{g(a+\xi\gamma+\xi'\gamma')}(-1)^{h(b+\xi\delta+\xi'\delta')}(-1)^{gh(1+\xi+\xi')}\\
&={1\over 2R_9}\sum_{k_9,\tilde l_9}e^{-{\pi \over (2R_9)^2\tau_2}|\tilde l_9+k_9\tau|^2} {1\over 2}\sum_{h,g}(-1)^{gk_9+h\tilde l_9}(-1)^{g(a+\xi\gamma+\xi'\gamma')}(-1)^{h(b+\xi\delta+\xi'\delta')}(-1)^{gh(1+\xi+\xi')}.\nonumber 
\end{align}
In the limit $R_9\to 0$, the only surviving contribution is for $(k_9,\tilde l_9)=(0,0)$, which yield
\be
 {1\over 2R_9}\sum_h\delta_{a+\xi \gamma+\xi'\gamma'+h(1+\xi+\xi'),0\, \rm mod \, 2}(-1)^{h(b+\xi\delta+\xi'\delta')}\equiv {1\over 2R_9} \Phi_{\xi\xi'}\!\big[{}^{a,\gamma,\gamma'}_{b,\, \delta,\, \delta'}\big] ,
\ee
where $\Phi_{\xi\xi'}$ satisfies
\be
\begin{aligned}
&\Phi_{10}\big[{}^{a,\gamma,\gamma'}_{b,\, \delta,\, \delta'}\big]=2 \delta_{a\gamma}\delta_{b\delta},\\
&\Phi_{\xi\xi}\big[{}^{a,\gamma,\gamma'}_{b,\, \delta,\, \delta'}\big]= (-1)^{ab}(-1)^{\xi[a(\delta+\delta')+b(\gamma+\gamma')+(\delta+\delta')(\gamma+\gamma')]}.
\end{aligned}
\ee
Hence, the total energy in the universe is better understood from a $d+1$-dimensional point of view, since 
\be
\int d^dx\, \V^{\rm o}\underset{R_9\to 0}{\sim}\int d^dx\,  2\pi{1\over 2R_9} \, \hat \V^{\rm o}\equiv \int d^{d+1}x\, \hat \V^{\rm o}, 
\ee
where we have defined 
\begin{eqnarray}
\hat \V^{\rm o}=\hat \V_0+\hat \V_*(R_{d-1})-{\hat \lambda_{\rm tw}\over R_{d-1}}.
\end{eqnarray}
In our notations, the first  contribution in $\hat \V^{\rm o}$ is independent of $R_{d-1}$, while the second does not,
\be
\begin{aligned}
\label{LVh}
\hat \V_0 &= -{1\over 2(2\pi)^{d+1}}\int_F {d\tau_1d\tau_2\over \tau_2^{1+{d+1\over 2}}} \, \Gamma_{(3,3)}\Gamma_{(6-d,6-d)}\, {1\over \eta^8\bar \eta^8}\, \hat Z,\\
\hat \V_*& = -{1\over 2(2\pi)^{d+1}}\int_{-{1\over 2}}^{1\over 2}d\tau_1 \int_0^{+\infty} {d\tau_2\over \tau_2^{1+{d+1\over 2}}} \sum_{\tilde m_{d-1}\neq 0}e^{-{\pi R_{d-1}^2\over \tau_2}\tilde m_{d-1}^2}\,\Gamma_{(3,3)}\Gamma_{(6-d,6-d)}\, {1\over \eta^8\bar \eta^8}\, \hat Z,
\end{aligned}
\ee
where we have denoted
\be
\hat Z =  {1\over 2}\sum_{a,b}(-1)^{a+b+ab}{\theta[{}^a_b]^4\over \eta^{4}} \, {1\over 2}\sum_{\gamma,\delta}{\bar \theta\big[{}^\gamma_\delta\big]^8\over \bar \eta^{8}}\, {1\over 2}\sum_{\gamma',\delta'}{\bar \theta\big[{}^{\gamma'}_{\delta'}\big]^8\over\bar \eta^{8}}\,\Phi_{\xi\xi'}\!\big[{}^{a,\gamma,\gamma'}_{b,\, \delta,\, \delta'}\big].
\ee
Notice that without $\Phi_{\xi\xi'}$ included, the above conformal block appears in the partition function of the parent supersymmetric $E_8\times E_8$ heterotic string. 
Finally, the third contribution involves 
\be
\begin{aligned}
\!\hat \lambda_{\rm tw}=-{1\over 2(2\pi)^{d+1}}&\int_F {d\tau_1d\tau_2\over \tau_2^{1+{d\over 2}}}\,\Gamma_{(6-d,6-d)}\\
& {1\over 2}\sum_{a,b}(-1)^{a+b+ab}  \,{1\over 2}\sum_{\gamma,\delta}\, {1\over 2}\sum_{\gamma',\delta'} \sum_{(H,G)\neq (0,0)}\!\!\!\omega\big[{}^{H,a,\gamma,\gamma'}_{G,\, b,\, \delta,\, \delta'}\big] \Phi_{\xi\xi'}\!\big[{}^{a,\gamma,\gamma'}_{b,\, \delta,\, \delta'}\big].
\end{aligned}
\ee

To describe the particle content of the conformal block $\hat Z$ which appears in $\hat \V_0+\hat \V_*$, we  introduce the $O(2n)$ affine characters, 
\begin{align}
O_{2n}&={\theta\big[{}^0_0\big]^n+\theta\big[{}^0_1\big]^n\over 2\eta^n}\, , &V_{2n}&={\theta\big[{}^0_0\big]^n-\theta\big[{}^0_1\big]^n\over 2\eta^n}\, ,\nonumber \\
S_{2n}&={\theta\big[{}^1_0\big]^n+(-i)^n\theta\big[{}^1_1\big]^n\over 2\eta^n}\, ,
&C_{2n}&={\theta\big[{}^1_0\big]^n-(-i)^n\theta\big[{}^1_1\big]^n\over 2\eta^n}\, .
\label{charac}
\end{align}

\ni $\bullet$ For $\xi=\xi'=0$, the only effect of $\Phi_{00}$ is to reverse spacetime chirality, $S_8\leftrightarrow C_8$, so that
\be
\hat Z = (V_8-C_8)(\bar O_{16}+\bar S_{16})(\bar O'_{16}+\bar S'_{16}),
\ee
where ``primed'' characters refer to those arising from the sums over $\gamma',\delta'$. 
In fact, this shows that  the model based on background~(\ref{back}) is supersymmetric in both T-dual decompactifications limits, $R_9\to +\infty$ and $1/(2R_9)\to +\infty$, with $E_8\times E'_8$ gauge symmetry.  While the light gravitinos present at large $R_9$ become infinitely massive and decouple in the $2R_9\to 0$ limit, other gravitinos arising from states winding $S^1(R_9)$ are becoming light.  As a result, this case does not yield the hard breaking of supersymmetry we are looking for. 

\ni $\bullet$ For $\xi=1, \xi'=0$, the presence of $\Phi_{10}$ breaks supersymmetry and preserves an $SO(16)\times E_8'$ gauge symmetry, 
\be
\hat Z = (O_8\bar V_{16}+V_8\bar O_{16}-S_8\bar S_{16}-C_8\bar C_{16})(\bar O'_{16}+\bar S'_{16}).
\ee
However, it also has a dramatic consequence, due to the following sector which contains a level-matched tachyon, 
\be
{O_8\over \eta^8}{\bar V_{16}\over \bar \eta^4}{\bar O'_{16}\over \bar \eta^4}={1\over q^{1\over 2}\bar q^{1\over 2}}+\cdots.
\ee
As a result, the model does not define a true vacuum, and the integral form of the energy density $\hat \V^{\rm 0}$ is divergent, which does not meet our objectives. 

\ni $\bullet$ For $\xi=\xi'=1$, supersymmetry is broken and the gauge symmetry is $SO(16)\times SO(16)$, 
\be
\begin{aligned}
\hat Z &= V_8(\bar O_{16} \bar O'_{16}+\bar S_{16}\bar S'_{16})-S_8(\bar O_{16} \bar S'_{16}+\bar S_{16}\bar O'_{16})\\
&+O_8(\bar V_{16} \bar C'_{16}+\bar C_{16}\bar V'_{16})-C_8(\bar V_{16} \bar V'_{16}+\bar C_{16}\bar C'_{16}).
\end{aligned}
\ee
Moreover, the potentially tachyonic left- and right-moving characters $O_8$ and $\bar O_{16}$ being nowhere multiplied,  there are no level-matched tachyon and $\hat \V^{\rm o}$ is finite. Hence, we can proceed with this consistent model in $d+1$ dimensions, where supersymmetry is explicitly broken~\cite{o16}. 

It turns out that $\hat \V_0$ is of order 1 and positive when so are the internal moduli ($c=\O(1)$ in Eq.~(\ref{expand}))~\cite{SNSM2,o16}. Moreover, expanding in $q,\bar q$ the integrand of $\hat \V_*$, we find that when the compact (though external) direction $x^{d-1}$ is large, namely $c\gg 1/R_{d-1}$, the contributions of the KK towers of modes propagating along $x^{d-1}$ dominate over all other string states, which are heavier. To be specific, we find
\be
\label{hV}
\hat \V_* =(\nF-\nB)\,  {v_{d+1}\over R_{d-1}^{d+1}}+\O\!\Bigg({c^{d\over 2}\over R_{d-1}^{{d\over 2}+1}}\, e^{-2\pi cR_{d-1}}\Bigg), \quad \where \quad \nF-\nB=8(264-d_{\rm en}),
\ee
and where $v_d$ was defined in field theory in Eq.~(\ref{res1}). 
In the above expressions, $\nF$, $\nB$ denote  the numbers of massless fermionic and bosonic degrees of freedom, while $v_{d+1}$ captures the contributions of their KK towers of states propagating in the ($\Z_2$-modded) spacetime direction $x^{d-1}$.

 Notice that $\hat \V=\hat \V_0+\hat \V_*$ reproduces the field theory answer, which is  written  in Eq.~(\ref{vd}) for a single bosonic degree of freedom, or in Eq.~(\ref{res1}) for a single bosonic or fermionic degree of freedom, but with three differences:
 
 $\bullet$ The string theory answer contains exponentially suppressed contributions arising from the heavy string modes which are not included in field theory. 
 
 $\bullet$ The string model with hard breaking we have constructed is in $d+1$ dimensions, rather than $d$ dimensions in field theory. Actually, starting from a $d$-dimensional model and then sending to zero the radius $R_9$  of an internal direction, we end up in string theory with  $d+1$ spacetime coordinates because of the reconstruction of an external direction from the towers of winding modes becoming massless. On the contrary, taking $R_9\to 0$ in field theory makes completely disappear an internal direction already invisible at small but finite $R_9$.  
 
$\bullet$  $\hat \V_0$ replaces (in $d+1$ dimensions) the divergent integral over the Euclidean momentum in Eq.~(\ref{vd}) (derived in second quantized formalism), or equivalently the term proportional to $\xi_d/R_9^d$ which diverges when $R_9\to 0$ in Eq.~(\ref{res1}) (derived in first quantized formalism). 

In the Casimir effect, as shown in Eqs~(\ref{finter}) and~(\ref{fexter}),  the forces per unit area exerted on the plate at $x^{d-1}=\pi R_{d-1}$ from the vacuum located on each sides are 
\be
\hat \F_{\rm int}=-{\partial (R_{d-1}\hat \V^{\rm o})\over \partial R_{d-1}}=-\hat \V_0-{\partial (R_{d-1}\hat \V_*)\over \partial R_{d-1}},\qquad \hat \F_{\rm ext}=\hat \V_0,
\ee
and they add to give the final answer
\be
\hat \F\equiv \hat \F_{\rm int}+\hat \F_{\rm ext}=(\nF-\nB)\,  {v_{d+1} d\over R_{d-1}^{d+1}}+\O\!\Bigg({c^{{d\over 2}+1}\over R_{d-1}^{d\over 2}}\, e^{-2\pi cR_{d-1}}\Bigg).
\ee
This is the analogue of the field theory result, Eqs~(\ref{fsdof}),~(\ref{fsdof2}), where the forces per unit area acting on either side of the plate were infinite, albeit with finite sum.
Therefore, string theory provides a more rigorous derivation of the resulting Casimir force when supersymmetry is not present at all (explicit breaking), to the extent that the forces on either side of each plate are individually finite. 


\section{Conclusion} 
\label{conclus}

We have shown that at weak coupling, string theory and quantum field theory yield similar results for the net force exerted on  parallel plates on which all degrees of freedom in the bulk are imposed vanishing boundary conditions (Dirichlet boundary conditions). To be specific, the answers are equal, up to exponentially suppressed corrections arising in the string computation from the heavy string states, which are absent in field theory. The net force can be derived by subtracting the energies associated with the vacuum comprised between the two plates, when they are at finite or infinite distance. 

In configurations where supersymmetry is spontaneously broken \`a la Scherk-Schwarz, both frameworks yield a finite energy density between the two plates, whatever is their distance. There is no need to subtract energies associated with two configurations to obtain  consistent results. Moreover, if this fact turns out to remain true in string theory  when supersymmetry is explicitly broken, it is not valid anymore  in quantum field theory.  Therefore, it is in this case that the infinite spectrum of the string, with unbounded masses,  plays a major role, as it provides the necessary UV regularization responsible for an absolute energy to make sense. Thus, it is in this case that string theory provides a more robust derivation of the Casimir effect, as compared to the usual quantum field theory analysis. 

We have derived these results by using a non-standard orbifold action. Whereas $\Z_2$ twists of  four internal directions are very common in string theory for reducing the number of supersymmetries, we have considered a $\Z_2$ generator $\G$ that twists one external compact direction and three internal coordinates. Moreover, for the two hyperplanes of the external space fixed by $\G$ to behave as ``conductive plates'', we have implemented in a way consistent with modular invariance the  projection $(\boldsymbol{1}-\G)/2$ on the spectrum running in the vacuum-to-vacuum amplitude at one-loop, which is orthogonal to the usual projector.  


\section*{Acknowledgement}
 
We are grateful to  Guillaume Bossard for fruitful discussions. 
A.K. is supported by the GSRT under the EDEIL/67108600 and by CNRS. The work of H.P. is partially supported by the Royal Society International Cost Share Award. A.K. would like to thank the CPHT of Ecole Polytechnique for hospitality.


%
%
%




\begin{thebibliography}{99}
    
    
    
 \bibitem{rev} For a review and further references, 
 see, 
  M.~Bordag, U.~Mohideen and V.~M.~Mostepanenko,
  ``New developments in the Casimir effect,''
  Phys.\ Rept.\  {\bf 353} (2001) 1
  \arXivold{quant-ph/0106045}.

\bibitem{Lam0}
  S.~K.~Lamoreaux,
  ``The Casimir force: Background, experiments, and applications,''
  Rept.\ Prog.\ Phys.\  {\bf 68} (2005) 201.

\bibitem{Lam1}
  S.~K.~Lamoreaux,
  ``Demonstration of the Casimir force in the 0.6 to 6 micrometers range,''
  Phys.\ Rev.\ Lett.\  {\bf 78} (1997) 5;
   Erratum: [Phys.\ Rev.\ Lett.\  {\bf 81} (1998) 5475].

\bibitem{jaffe}
  R.~L.~Jaffe,
  ``The Casimir effect and the quantum vacuum,''
  Phys.\ Rev.\ D {\bf 72} (2005) 021301
  \arXivold{hep-th/0503158}.

\bibitem{o16}
  L.~Alvarez-Gaum\'e, P.~H.~Ginsparg, G.~W.~Moore and C.~Vafa,
  ``An $O(16) \times O(16)$ heterotic string,''
  Phys.\ Lett.\ B {\bf 171} (1986) 155.
  
\bibitem{kiritsis} 
 E.~Kiritsis,
  ``String theory in a nutshell,'' Princeton University Press, 2007


 \bibitem{IZ} For the spin-1 case, see C.~Itzykson and J.~B.~Zuber,
  ``Quantum field theory,''
  New York, Usa: McGraw-Hill (1980) 705 p. (International Series In Pure and Applied Physics)

 \bibitem{SS}
  J.~Scherk and J.~H.~Schwarz,
  ``Spontaneous breaking of supersymmetry through dimensional reduction,''
  Phys.\ Lett.\  B {\bf 82} (1979) 60.
%
  \bibitem{SS1}
  J.~Scherk and J.~H.~Schwarz,
 ``How to get masses from extra dimensions,''
  Nucl.\ Phys.\ B {\bf 153} (1979) 61.
  
  \bibitem{SSstring}
  R.~Rohm,
  ``Spontaneous supersymmetry breaking in supersymmetric string theories,''
  Nucl.\ Phys.\ B {\bf 237} (1984) 553.
  
  \bibitem{s1}
  C.~Kounnas and M.~Porrati,
  ``Spontaneous supersymmetry breaking in string theory,''
  Nucl.\ Phys.\ B {\bf 310} (1988) 355.
  
  \bibitem{s2}
  S.~Ferrara, C.~Kounnas and M.~Porrati,
  ``Superstring solutions with spontaneously broken four-dimensional supersymmetry,''
  Nucl.\ Phys.\  B {\bf 304} (1988) 500.
  
  \bibitem{s3}
  S.~Ferrara, C.~Kounnas, M.~Porrati and F.~Zwirner,
  ``Superstrings with spontaneously broken supersymmetry and their effective theories,''
  Nucl.\ Phys.\ B {\bf 318} (1989) 75.

  \bibitem{SSstring2}
  C.~Kounnas and B.~Rostand,
  ``Coordinate-dependent compactifications and discrete symmetries,''
  Nucl.\ Phys.\ B {\bf 341} (1990) 641.


  
\bibitem{crisol}
  T.~Catelin-Jullien, C.~Kounnas, H.~Partouche and N.~Toumbas,
  ``Thermal/quantum effects and induced superstring cosmologies,''
  Nucl.\ Phys.\ B {\bf 797} (2008) 137
  \arXivold{arXiv:0710.3895} [hep-th].

\bibitem{cr1}
  T.~Catelin-Jullien, C.~Kounnas, H.~Partouche and N.~Toumbas,
  ``Induced superstring cosmologies and moduli stabilization,''
  Nucl.\ Phys.\ B {\bf 820} (2009) 290
   \arXivold{arXiv:0901.0259} [hep-th].
  
  \bibitem{attract}
  F.~Bourliot, C.~Kounnas and H.~Partouche,
  ``Attraction to a radiation-like era in early superstring cosmologies,''
  Nucl.\ Phys.\ B {\bf 816} (2009) 227
   \arXivold{arXiv:0902.1892} [hep-th].
  
 \bibitem{B1}
  F.~Bourliot, J.~Estes, C.~Kounnas and H.~Partouche,
  ``Cosmological phases of the string thermal effective potential,''
  Nucl.\ Phys.\ B {\bf 830} (2010) 330
  \arXivold{arXiv:0908.1881} [hep-th].
  
  \bibitem{B2}
  J.~Estes, C.~Kounnas and H.~Partouche,
  ``Superstring cosmology for $\N_4 = 1 \to 0$ superstring vacua,''
  Fortsch.\ Phys.\  {\bf 59} (2011) 861
  \arXivold{arXiv:1003.0471} [hep-th].
  
    \bibitem{B3}
  L.~Heurtier, T.~Coudarchet and H.~Partouche, ``Spontaneous dark-matter mass generation along cosmological attractors in string theory,''   \arXivold{arXiv:1812.10134} [hep-th].


  
  \bibitem{SNSM}
  C.~Kounnas and H.~Partouche,
  ``Super no-scale models in string theory,''
  Nucl.\ Phys.\ B {\bf 913} (2016) 593
  \arXivold{arXiv:1607.01767} [hep-th].
  
 \bibitem{SNSM2}
    C.~Kounnas and H.~Partouche,
  ``$\mathcal N=2 \to 0$ super no-scale models and moduli quantum stability,''
  Nucl.\ Phys.\ B {\bf 919} (2017) 41
  \arXivold{arXiv:1701.00545} [hep-th].
  


\bibitem{McClain:1986id}
  B.~McClain and B.~D.~B.~Roth,
  ``Modular invariance for interacting bosonic strings at finite temperature,''
  Commun.\ Math.\ Phys.\  {\bf 111} (1987) 539.
  
\bibitem{M1}
  K.~H.~O'Brien and C.~I.~Tan,
  ``Modular invariance of thermopartition function and global phase structure
  of heterotic string,''
  Phys.\ Rev.\  D {\bf 36} (1987) 1184.

  


\end{thebibliography}
\end{document}